\documentclass[preprint]{elsarticle}
\usepackage{graphicx}
\usepackage{dcolumn}
\usepackage{amssymb}
\usepackage{url}
\usepackage{hyperref}

\def\units#1{~\hbox{$\,{\rm #1}$}}
\def\degrees{\hbox{$^\circ$}}

\journal{Astroparticle Physics}

\begin{document}

\begin{frontmatter}
\title{The DArk Matter Particle Explorer mission}
\author{DAMPE collaboration: J. Chang$^{a}$\footnote{Corresponding author (email: chang@pmo.ac.cn)}, G. Ambrosi$^{b}$, Q. An$^{c}$, R. Asfandiyarov$^{d}$, P. Azzarello$^{d}$, P. Bernardini$^{e,f}$,
B. Bertucci$^{g,b}$, M. S. Cai$^{a}$, M. Caragiulo$^{h}$, D. Y. Chen$^{a,i}$
% ...
}

\address[a]{Key Laboratory of Dark Matter and Space Astronomy, Purple Mountain Observatory, Chinese Academy of Sciences, Nanjing 210008, China}
\address[b]{Istituto Nazionale di Fisica Nucleare Sezione di Perugia, I-06123 Perugia, Italy}
\address[c]{State Key Laboratory of Particle Detection and Electronics, University of Science and Technology of China, Hefei 230026, China}
\address[d]{Department of Nuclear and Particle Physics, University of Geneva, CH-1211, Switzerland}
\address[e]{Universit\`a del Salento - Dipartimento di Matematica e Fisica "E. De Giorgi", I-73100, Lecce, Italy}
\address[f]{Istituto Nazionale di Fisica Nucleare (INFN) - Sezione di Lecce , I-73100 , Lecce, Italy}
\address[g]{Dipartimento di Fisica e Geologia, Universit\`a degli Studi di Perugia, I-06123 Perugia, Italy}
\address[h]{Istituto Nazionale di Fisica Nucleare Sezione di Bari, I-70125, Bari, Italy}
\address[i]{University of Chinese Academy of Sciences, Yuquan Road 19, Beijing 100049, China}
\address[j]{Institute of Modern Physics, Chinese Academy of Sciences, Nanchang Road 59, Lanzhou 730000, China}
\address[k]{National Space Science Center, Chinese Academy of Sciences, Nanertiao 1, Zhongguancun, Haidian district, Beijing 100190, China}
\address[l]{Institute of High Energy Physics, Chinese Academy of Sciences, YuquanLu 19B, Beijing 100049, China}
\address[m]{ASI Science Data Center (ASDC), I-00133 Roma, Italy}
\address[n]{Dipartimento di Fisica "M.Merlin" dell'Univerisity e del Politecnico di Bari, I-70126, Bari, Italy}
\address[o]{School of Astronomy and Space Science, University of Science and Technology of China, Hefei, Anhui 230026, China}

\begin{abstract}
The DArk Matter Particle Explorer (DAMPE), one of the four scientific space science missions
 within the framework of the Strategic Pioneer Program on Space Science of the Chinese Academy
 of Sciences, is a general purpose high energy cosmic-ray and gamma-ray observatory, which was
 successfully launched on December 17th, 2015 from the Jiuquan Satellite Launch Center.
 The DAMPE scientific objectives include the study of galactic cosmic rays up to $\sim 10$ TeV and hundreds of TeV
 for electrons/gammas and nuclei respectively, and the search for dark matter signatures in their spectra.
 In this paper we illustrate the layout of the DAMPE instrument, and discuss the results of beam tests and calibrations
 performed on ground. Finally we present the expected performance in space and give an overview
 of the mission key scientific goals.
\end{abstract}
\end{frontmatter}

\section{introduction}
The interest in space-borne particle/astroparticle physics experiments is growing. The achievements of the early space-borne
particle detectors such as IMP \cite{IMP}, HEAO-3 \cite{HEAO}, ACE \cite{ACE} lead to more advanced experiments, namely EGRET \cite{EGRET} , AMS-01 \cite{AMS01}, PAMELA \cite{PAMELA}, AGILE \cite{AGILE}, Fermi \cite{FERMI}, AMS-02 \cite{AMS02}
and CALET \cite{CALET}. Additionally, there have been many balloon and ground based experiments including BESS \cite{BESS},
 IMAX \cite{IMAX}, HEAT \cite{HEAT}, ATIC \cite{ATIC}, CAPRICE \cite{CAPRICE}, CREAM \cite{CREAM}, WIZARD \cite{WIZARD},
 Fly's Eye \cite{FLY}, H.E.S.S \cite{HESS}, MAGIC \cite{MAGIC}, ARGO-YBJ experiment \cite{ARGO2016b}, VERITAS \cite{VERITAS}, Pierre Auger Observatory \cite{AUGER}, HAWC \cite{HAWC} etc. Our understanding of the high-energy universe has been revolutionized thanks to the successful operation of these experiments.

The DArk Matter Particle Explorer (DAMPE~\cite{Chang2014}), initially named TANSUO~\cite{Chang2009,WuJ2011,ChangFan2011},
was successfully launched into a sun-synchronous orbit at the altitude
of 500 km on 2015 December $17^{\rm th}$ from the Jiuquan launch base. DAMPE
offers a new opportunity for advancing our knowledge of cosmic rays, dark matter, and gamma-ray astronomy. In this paper a detailed overview of the DAMPE instrument is provided, the expected instrumental performance based on extensive GEANT4 simulations are presented, and the key scientific objectives are outlined and discussed.



DAMPE is able to detect electrons/positrons, gamma rays, protons, helium nuclei and other
heavy ions in a wide energy range with much improved energy resolution and large acceptance (see Table~\ref{tab-InstSumm} for summary of the instrument parameters). The primary observing mode is the sky survey in a sun-synchronous orbit at the
altitude of 500 km, and it is expected to cover the full sky at least four times in two years. The main scientific objectives addressed by DAMPE include: (1) understanding the mechanisms of particle acceleration operating
in astrophysical sources, and the propagation of cosmic rays in the the Milky Way; (2) probing the nature of dark matter; and (3) studying the gamma-ray emission from Galactic and extragalactic sources.

\begin{tiny}
\begin{table}[!h]
\begin{center}
\caption{Summary of the design parameters and expected performance of DAMPE instrument}\label{tab-InstSumm}
\begin{tabular}{llllll}
\hline
Parameter	                                 &	Value                     \\
\hline
Energy range of $\gamma$-rays/electrons	     & ~~~5 GeV$-$10 TeV 			        \\
Energy resolution$^a$ of $\gamma$-rays/electrons & $\leq 1.5\%$ at 800 GeV              \\
Energy range of protons/heavy nuclei	     & ~~~50 GeV$-$100 TeV		            \\
Energy resolution$^a$ of protons                 & $\leq 40\%$ at 800 GeV               \\
Effective area at normal incidence ($\gamma$-rays)     & $1100~{\rm cm}^{2}$ at 100 GeV          \\
Geometric factor for electrons & $0.3~{\rm m^{2}~sr}$ above 30 GeV    \\
Photon angular resolution$^{b}$	 & $\leq 0.2\degrees$ at 100 GeV	        \\
Field of View (FoV)                          & $\sim$1.0 sr                         \\
\hline
\end{tabular}
\end{center}
Notes: $^a$$\sigma_E/E$ assuming Gaussian distribution of energies.
$^b$The 68\% containment radius.
\end{table}
\end{tiny}

\begin{figure}[!ht]
\includegraphics[scale=0.5]{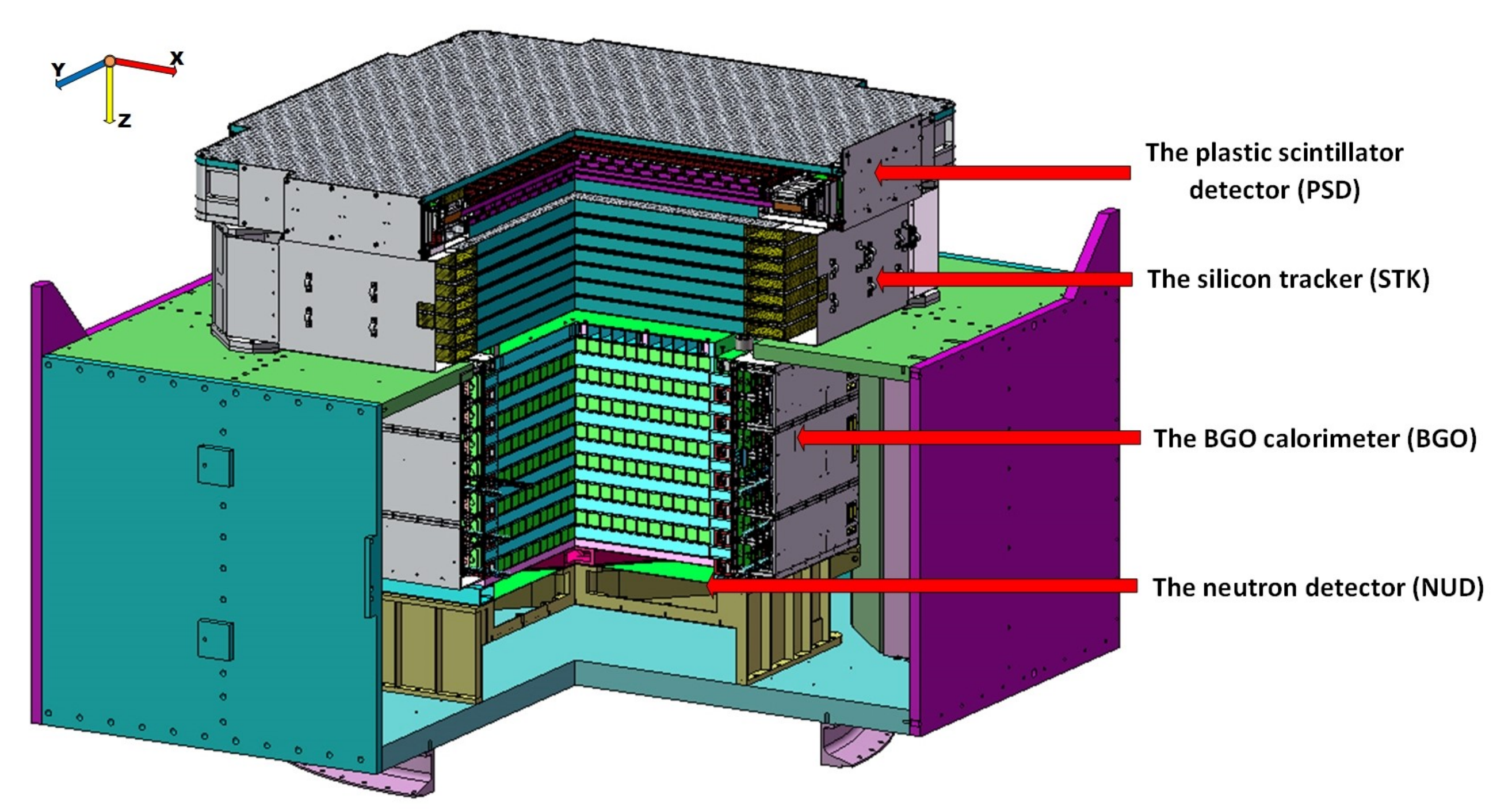}
\caption{Schematic view of the DAMPE detector.} \label{fig:SideView}
\end{figure}

\section{The DAMPE instrument}
Fig.\ref{fig:SideView} shows a schematic view of the DAMPE detector. It consists of a Plastic Scintillator strip Detector (PSD), a Silicon-Tungsten
tracKer-converter (STK), a BGO imaging calorimeter and a NeUtron Detector (NUD). The PSD provides charged-particle background rejection for gamma
rays (anti-coincidence detector) and measures the charge of incident particles; the STK measures the charges and the trajectories of charged particles,
and allows to reconstruct the directions of incident photons converting into $e^{+} e^{-}$ pairs; the hodoscopic BGO calorimeter, with a total depth of
about 32 radiation lengths, allows to measure the energy of incident particles with high resolution and to provide efficient electron/hadron identification;
finally, the NUD provides a independent measurement and further improvement of the electron/hadron identification.

\subsection{The Plastic Scintillation array Detector (PSD)}

The main purpose of the PSD is to provide charged-particle background rejection for the gamma ray detection and to measure the absolute
value of the charge (hereafter $Z$) of incident high-energy particles in a wide range (i.e., $Z \leq 26$). Therefore high detection efficiency, large dynamic range, and good charge resolution are required for charged particle detection of PSD. The main instrumental parameters of the PSD are summarized in Table~\ref{tab:ParaPSD}.

\begin{tiny}
\begin{table}[!ht]
\begin{center}
\caption{Summary of the designed parameters and expected performance of PSD.} \label{tab:ParaPSD}
\begin{tabular}{llllll}
\hline
Parameter	      &  Value                 \\
\hline
Active area	      &  $\geq$82 cm $\times$ 82 cm  \\
Number of layers      &  2              \\
Dynamic range	      &  Electrons, ions ($Z\leq$26)		        \\
Charge resolution$^a$ &  $\leq25\%$ for $Z = 1$              \\
Detector efficiency of single module   	&  $\geq0.95$ for MIPs                                    \\
Position resolution$^b$	& $\leq$2 cm				\\
\hline
\end{tabular}
\end{center}
Note: $^a$$\sigma_Z/Z$ assuming Gaussian distribution.
$^b$Geometry size of the PSD bar.
\end{table}
\end{tiny}

\begin{figure}[!ht]
	\centering
	\includegraphics[scale=0.4]{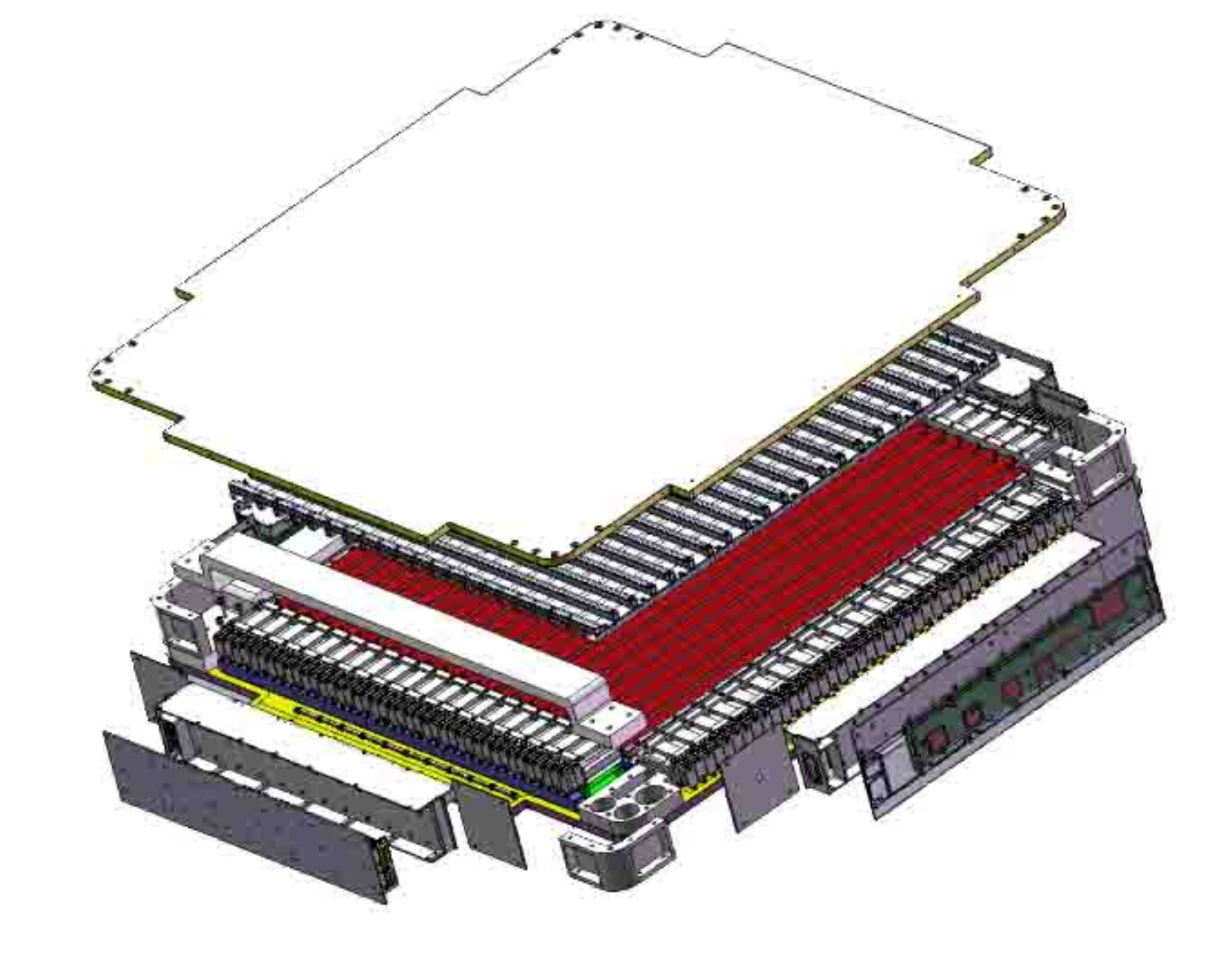}
	\caption{Schematic view of the PSD.} \label{fig:PSD}	
\end{figure}

A schematic view of the PSD is shown in Fig.~\ref{fig:PSD}. The PSD has an active area of $82.5 \times 82.5~{\rm cm^2}$, that is larger than the on-axis cross section of other sub-detectors of DAMPE~\cite{YuYH2017}.
The PSD consists of 82 plastic scintillator (EJ-200 produced by Eljen~\cite{eljen}) bars arranged in two planes, each with a double layer
configuration. Each bar is 88.4 cm long with a $2.8~{\rm cm} \times 1.0~{\rm cm}$ cross section; the signals are readout by two Hamamatsu R4443 Photomultiplier Tubes (PMTs)
coupled to the ends of each scintillator bar. The bars in the top plane are perpendicular to those in the bottom plane.
The bars of the two layers of a plane are staggered by 0.8 cm, allowing a full coverage of the detector with the active area of scintillators without any gap. As the efficiency of a single layer is $\geq 0.95$, the PSD provides an overall efficiency $\geq 0.9975$ for charged particles.
The segmented structure of the PSD allows to suppress the spurious veto signals due to the  ``backsplash effect", which can lead to a misidentification of
gamma rays as charged particles. This phenomenon was observed in EGRET and was found to be significant for photon energies in the GeV region and above. A similar choice of the segmented design
was adopted in the AGILE~\cite{AGILE} and the Large Area Telescope onboard the Fermi telescope (Fermi-LAT)~\cite{FERMI}, both equipped with anti-coincidence detectors consisting of
plastic scintillator tiles.

Since the PSD is used to identify cosmic-ray nuclei from helium to iron ($Z=26$), a wide dynamic range extending up to $\sim 1400$
times the energy deposition of a minimum ionizing particle (MIP)~\footnote{A singly charged MIP at normal incidence, which is assumed as
reference, deposits on average about 2 MeV in a single PSD bar.}  is required. To cover such a broad range with good energy resolution, a double dynode readout scheme
for each PMT has been implemented. Signals from the dynode with high gain cover the range from 0.1 MIPs to 40 MIPs, while those from the dynode with
low gain cover the range from 4 MIPs to 1600 MIPs; the overlap region can be used for cross calibration~\cite{YuYH2017,ZhouY2016a,ZhouY2016b}.

The dynode signals are coupled to VA160 ASIC chip developed by IDEAS~\cite{IDEAS}. This chip integrates the charge sensitive preamplifier, the shaper
and the holding circuit for 32 channels. Four groups of front-end electronics (FEE) chips are placed at all the sides of the PSD, and each FEE processes
82 signal channels from 41 PMTs in each side. With each group of FEE, there is also a high-voltage fan-out board, which supplies the high-voltages to all
the 41 PMTs in the same side.

The detector plane, the four groups of FEEs, and the high-voltage fan-out boards are mounted together with the mechanical support. To minimize
the materials used in the active area, this
mechanical support is mainly made by honeycomb boards with Carbon Fiber Reinforced Plastics (CFRP) as the skin (see Fig.\ref{fig:PSD}). Due to the large difference of temperature coefficients between the plastic scintillator and the CFRP, in order to avoid the damage due to large temperature variations, each of the detector modules is only fixed on the support with one end, and the other end is only constrained by a U-shape clamp while keeping the moving freedom along the bar direction.

\begin{figure}[!ht]
	\centering
	\includegraphics[scale=0.5]{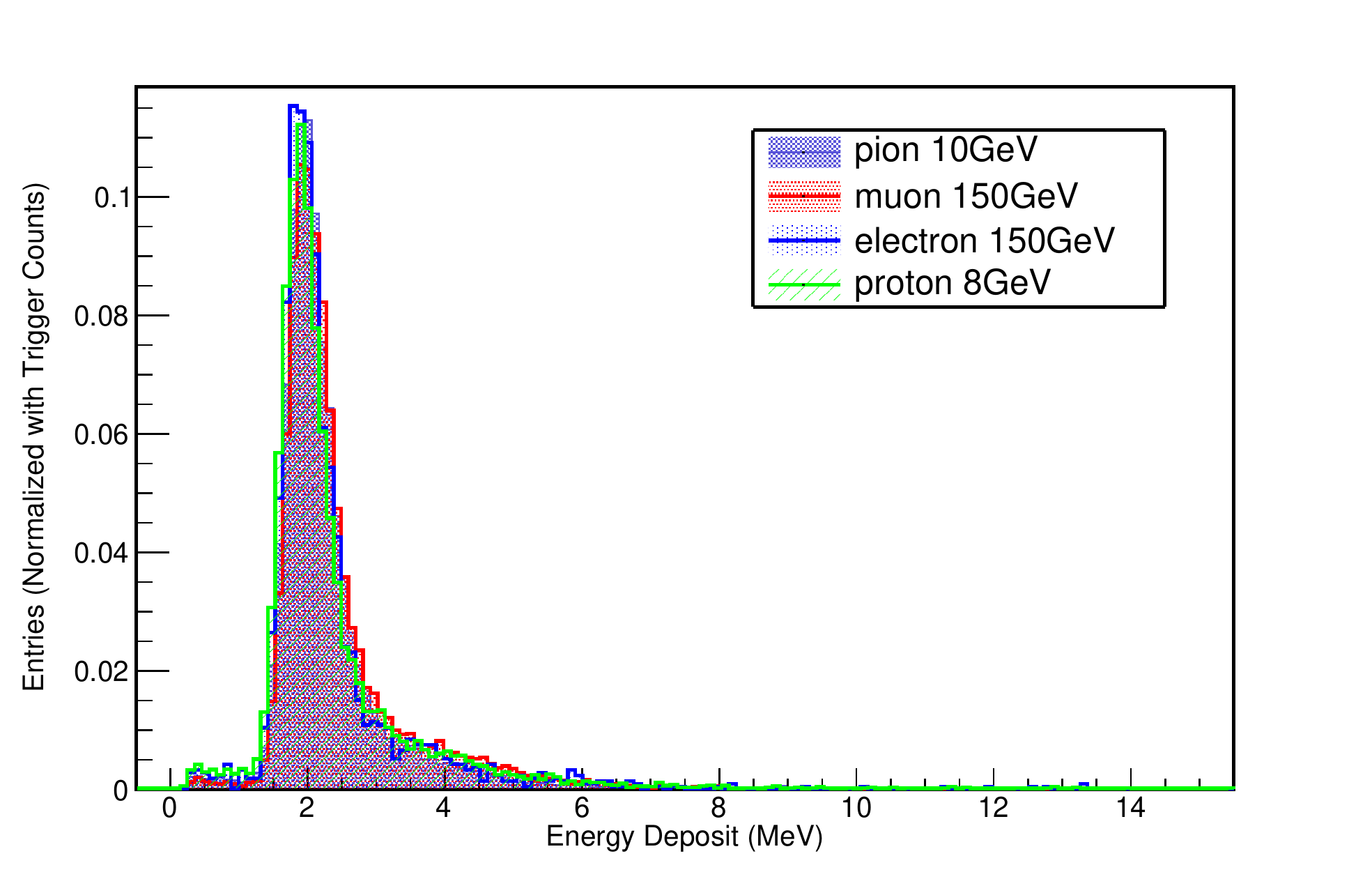}
	\caption{Energy deposited in the PSD as measured on beam tests for different species of $Z=1$ particles.} \label{fig:PSD_Z1}
\end{figure}

In 2014 and 2015, the Engineering Qualification Model (EQM) of DAMPE has been extensively tested on different particle beams, namely high energy gamma-rays
($0.5-150$ GeV), electrons ($0.5-250$ GeV), protons ($3.5-400$ GeV), $\pi^{-}$ ($3-10$ GeV), $\pi^{+}$ ($10-100$ GeV), muons ($150$ GeV) and various nuclei produced by
fragmentation of Argon ($30-75$ GeV/n) and Lead (30 GeV/n) in the European Organization for Nuclear Research (CERN).

Fig.~\ref{fig:PSD_Z1} shows the energy deposited in the PSD for different species of charged particles with $Z=1$.
We find that the peaks can be well described by Landau distribution due to the limited number
of photons collected by the PMTs. Despite their very different mass and energy, the energy deposits for leptons (electrons, muons) and hadrons
(pions, protons) are nearly the same.
%
%
For a singly charged incident particle, the energy resolution is $\sim$10$\%$ which can be regarded as the charge resolution of PSD.

\begin{figure}[!ht]
	\centering
	\includegraphics[scale=0.62]{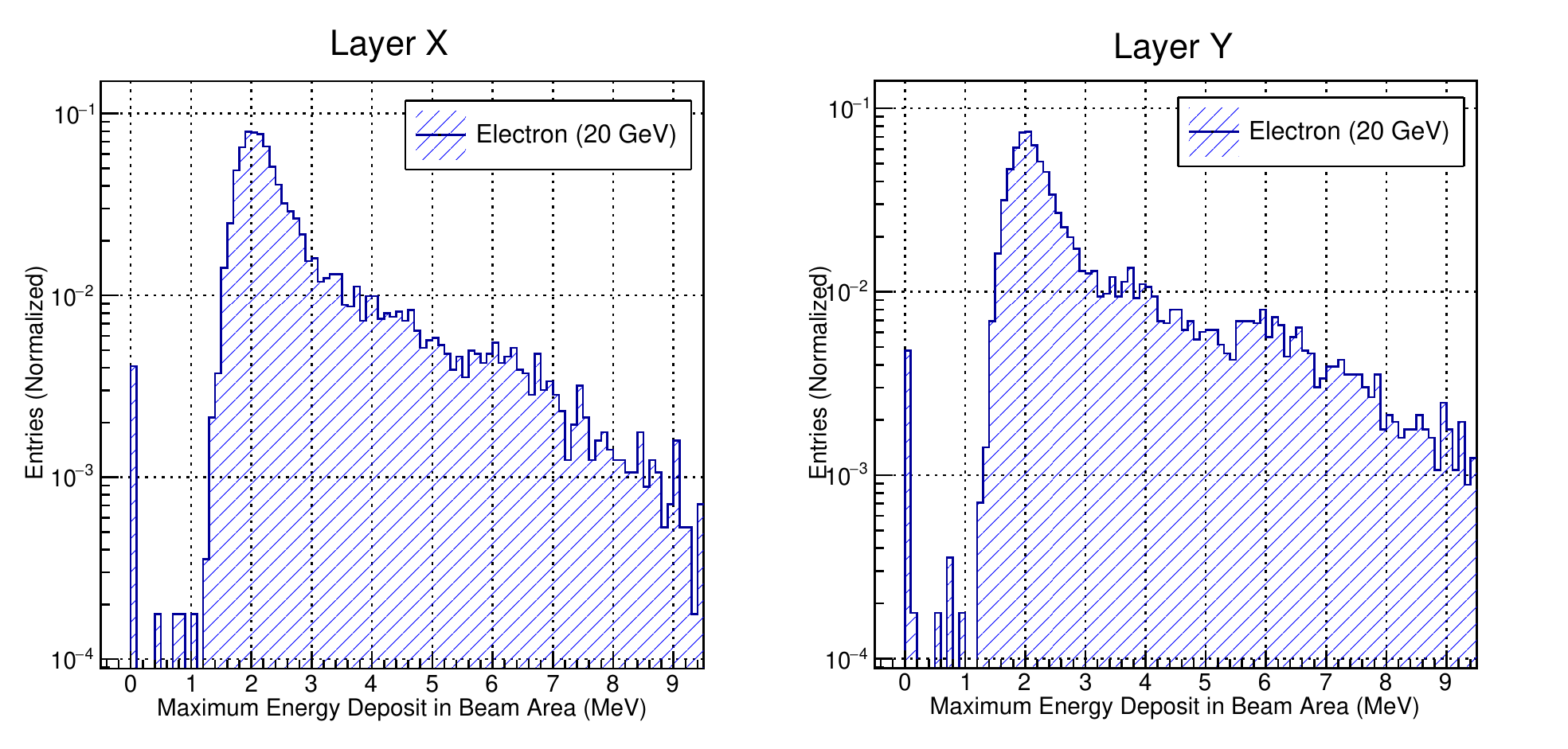}
	\caption{Energy deposit for 20 GeV electrons in the PSD modules lying in the beam spot region (see text).} \label{fig:PSD_electron}	
\end{figure}

As mentioned above, in order to effectively separate gamma rays from charged particles, the PSD should have a high detection efficiency for $Z=1$ particles.
Such a performance was checked with electron beams of different energies. Fig.~\ref{fig:PSD_electron} shows the spectra of deposited energy of 20 GeV electron beam in both X and Y layers. To minimize the influence of the backsplash effect, only modules within the beam spot area have been considered.
By setting the threshold at 1 MeV, which corresponds to about 0.5 MIP, an efficiency higher than 0.994 has been achieved for each layer.

\begin{figure}[!ht]
	\centering
	\includegraphics[scale=0.5]{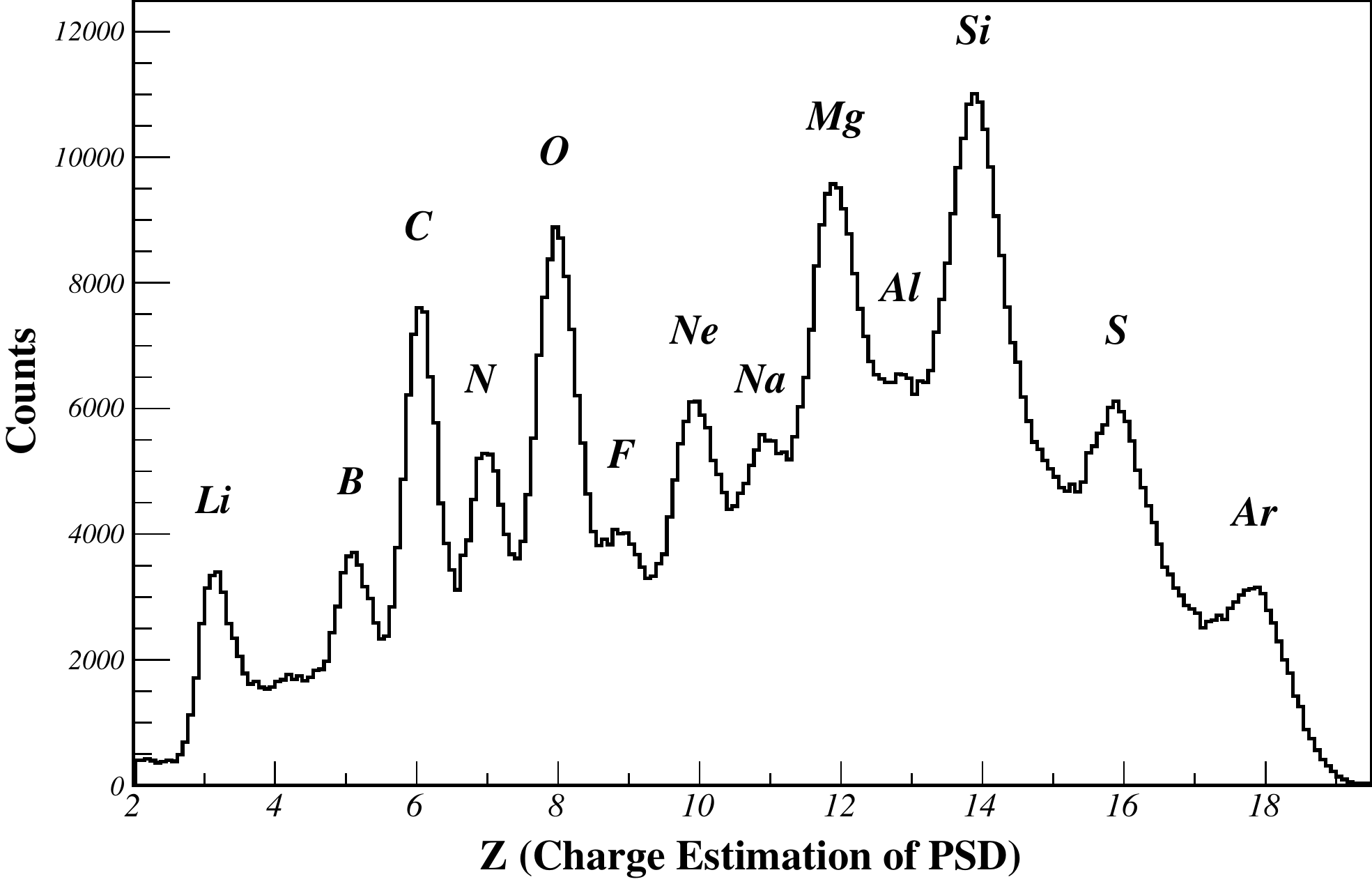}
	\caption{Reconstructed charge spectra of PSD for nuclei with $A/Z =2$, generated by a 40 GeV/n $^{40}$Ar beam.
	         The helium peak has been removed for clarity.} \label{fig:PSD_ion}	
\end{figure}

The performance of the PSD has been also tested with the relativistic heavy ion beams at CERN.
In this test, the primary Argon beam of $40$ GeV/n was sent onto a 40 mm polyethylene target, and  the secondary fragments with $A/Z=2$ were selected by beam magnets,
thus allowing to study the PSD response to all the stable nuclei with $Z=2\div18$. Fig.~\ref{fig:PSD_ion} shows the reconstructed charge spectra for different ions
($Z>2$) from one PSD module within the beam spot. In this figure the Helium contribution
has been removed for clarity (the He fraction is much higher than that of other ion species).
The signals from both sides of each module are used (geometric mean) and the quenching effect has been corrected based on the ion response
from the same test.

It can be seen that all the elements from Lithium $(Z=3)$ to Argon $(Z=18)$ can be identified clearly.
By applying a multi-Gaussian fit to the spectrum, we get the charge resolution of PSD for all ion species with the typical value of 0.21 for Helium and 0.48 for Argon. The charge resolution is expected to be better in space, because of much lower ion rates with respect to the case of beam tests. The results show that the position of the Ar peak in the raw Analog-Digital Conversion (ADC) spectrum
for different PSD modules is only $\sim20\%$ of the full dynamic range.
By simple extrapolation using the Birks-Chou law~\cite{Dwyer85}, this validates that the PSD can cover ion species up to Iron $(Z=26)$.

\subsection{The Silicon-Tungsten tracKer-converter (STK)}

\begin{figure}[!ht]
\begin{center}
\includegraphics[width=0.7\textwidth]{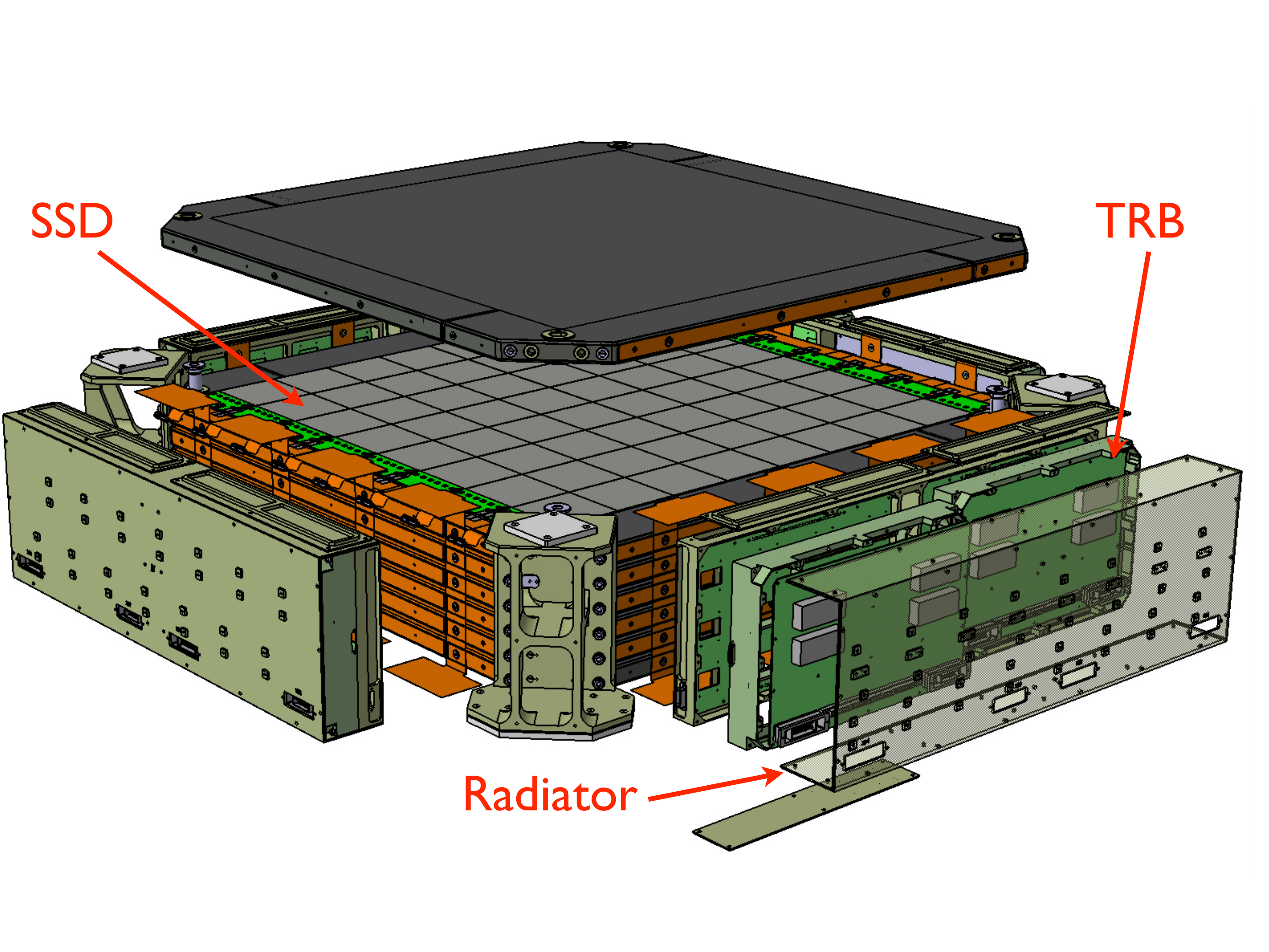}
\vskip -0.8cm
\caption{  Exploded view of the STK.}
\label{fig_stk}
\end{center}
\end{figure}

The DAMPE STK is designed to accomplish the
following tasks: precise particle track reconstruction with a
resolution better than 80 $\mu$m for most of the incident angles,
measurement of the electrical charge of incoming cosmic rays, and photon
conversion to electron-positron pairs \cite{wu_proceeding,
azzarello_proceeding}. The DAMPE tracker-converter system combines the main features of the
previous successful missions including AGILE~\cite{AGILE}, Fermi-LAT~\cite{FERMI} and AMS-02~\cite{AMS02}. It is composed of six position-sensitive
 double (X and Y) planes of silicon detectors with a total area of about 7
m$^2$, comparable with the total silicon surface of the AMS-02 tracker. Multiple thin tungsten layers have been inserted in the tracker structure in order to enhance the photon conversion rate while keeping negligible multiple scattering of electron/positron pairs (above $\sim 5\,$GeV). The total thickness of STK corresponds to about one radiation length, mainly due to
the tungsten layers.  An exploded view of the STK is shown in Fig.~\ref{fig_stk}, and a summary of the DAMPE STK instrument parameters is given in
Table~\ref{tab:ParaSTK}.

\begin{tiny}
\begin{table}[!ht]
\begin{center}
\caption{Summary of designed parameters of STK.} \label{tab:ParaSTK}
\begin{tabular}{llllll}
\hline
Parameter                                &    Value                 \\
\hline
    Active area of silicon detectors       &   0.55$\,$m$^2$ $\times$ 12 layers \\
    Thickness of each silicon layer             &   320 $\mu$m             \\
    Silicon strip pitch                           &   121 $\mu$m                \\
    Thickness of tungsten layers                           &   3 $\times$ 1 mm              \\
    Total radiation length                      &  0.976 (X$_0$)                                    \\
    Spatial resolution$^a$                   & $<80\mu$m within $60^\circ$ incidence \\
    Power consumption  &  82.7$\,$W      \\
    Total mass         &  154.8$\,$kg    \\
\hline
\end{tabular}
\end{center}
Note: $^a$68\% extension range.
\end{table}
\end{tiny}

The mechanical structure is made of 7 supporting trays of aluminum
honeycomb layers sandwiched between two CFRP face sheets of \mbox{0.6 mm} thick. The second, third and fourth
planes are equipped with 1 mm thick tungsten plates glued into the CFRP
sheet inside the tray, which was produced by Composite Design
S\`arl~\cite{cite:CompositeDesign}. The overall structure is light but stable in order
to withstand the vibrations and accelerations during the launch. The alignment of each tungsten plate
with respect to the 4 corners of the tray has been checked with a X-ray
scan at CERN.

\begin{figure}[htbp]
\begin{center}
\includegraphics[width=1.0\textwidth]{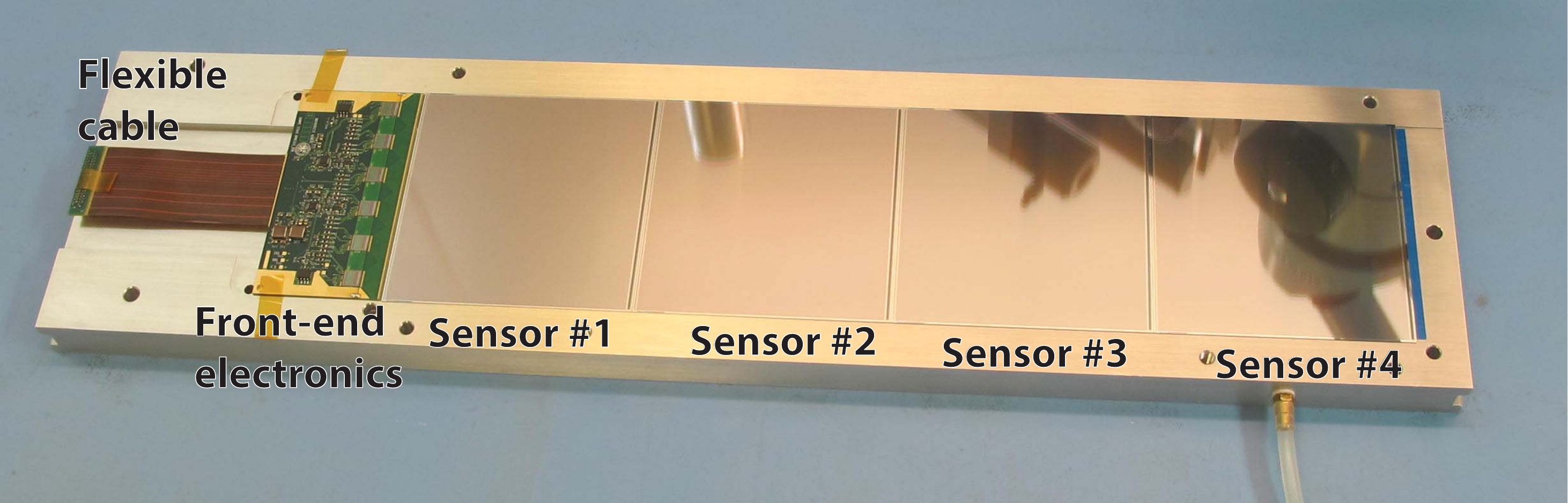}
\caption{ The STK single ladder, made by four SSDs.}
\label{fig_stk_ladder}
\end{center}
\end{figure}

The STK detector is equipped with a total of 768 single-sided AC-coupled
silicon micro-strip detectors (SSD). Four SSDs are assembled together with
a wire bonded strip-to-strip connection to form a silicon detector ladder, shown in Fig.~\ref{fig_stk_ladder}. The total strip
length along a ladder is about 37 cm. The ladders are glued on the seven support
trays to form the 12 STK silicon layers. Each silicon layer consists of
16 ladders, as shown in Fig.~\ref{fig_stk}. The two sides of the
five central trays are both equipped with 16 ladders each, while for the
top and the bottom plane only one side is equipped with the silicon
ladders. All the planes are piled up together to form the full tracker system. The
silicon ladders on the bottom surface of each tray are placed
orthogonal with respect to the ones of the top surface of the lower
tray, in order to measure the X-Y coordinates of the incident particles.
The inter-distance between two consecutive silicon layers is \mbox{$\sim 3$ mm}.

\begin{figure}[htbp]
\begin{center}
\begin{tabular}{ll}
 \includegraphics[width=0.5\textwidth]{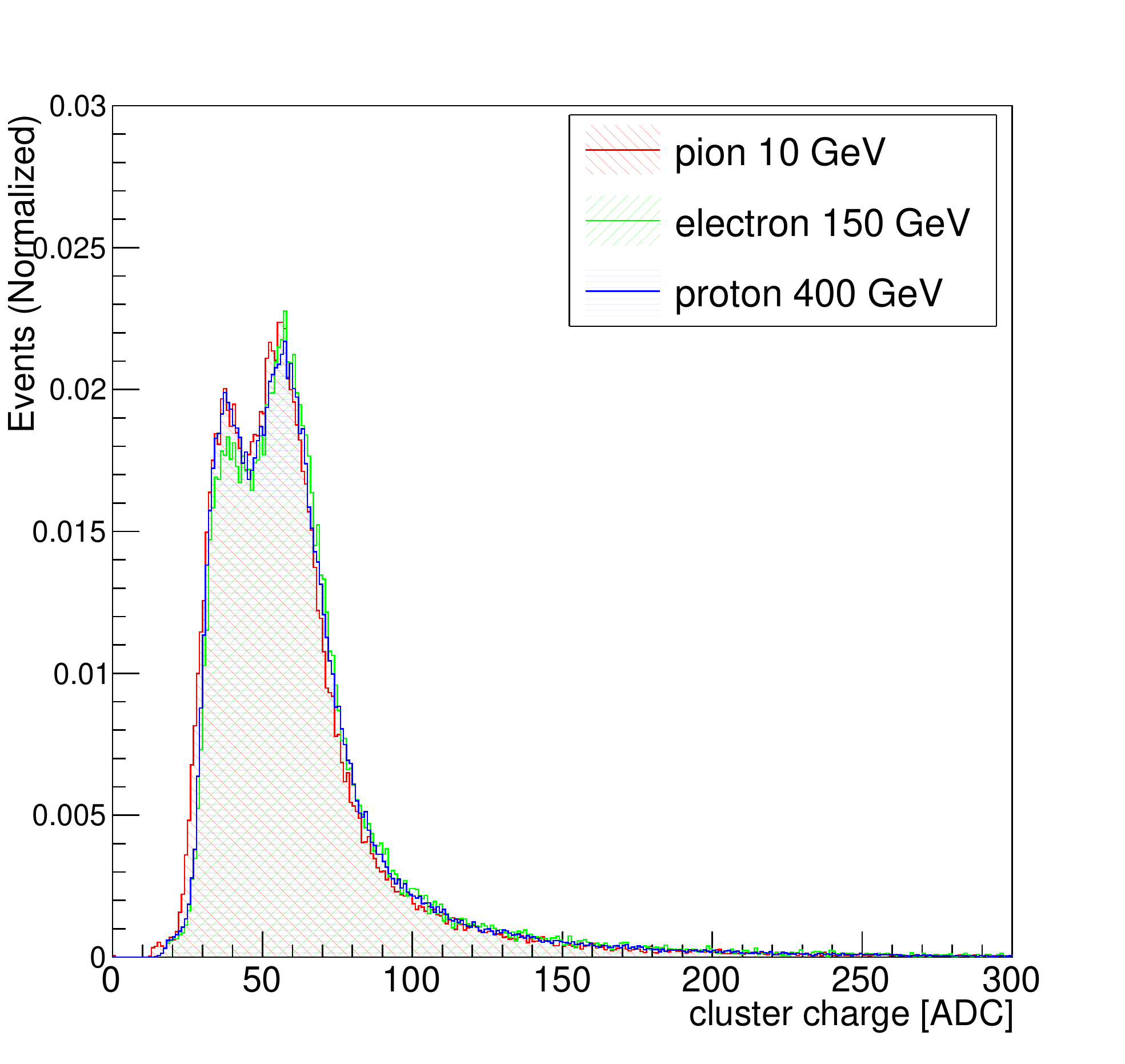}  &
 \includegraphics[width=0.5\textwidth]{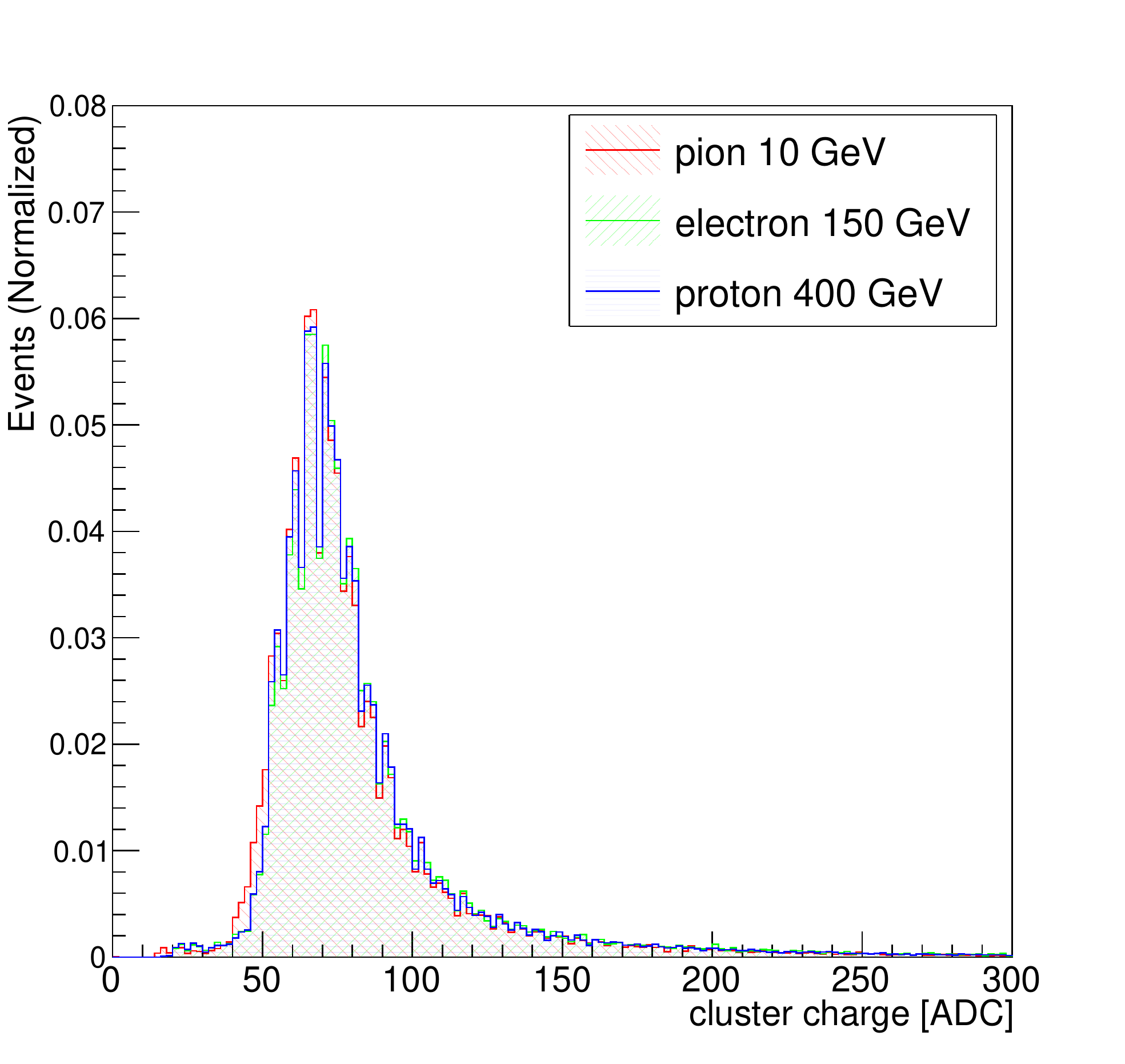} \\
\end{tabular}{}
\caption{{\it  (Left)} STK cluster charge response in terms of ADC counts for different singly charged particles with an incident angle of 0$^\circ$ (i.e.  the particle beam is orthogonal with respect to the silicon layers). {\it (Right)} Cluster charge distributions after the correction. All distributions are normalized to unit for shape comparison.}
\label{fig_cluster_charge_response}
\end{center}
\end{figure}

The silicon micro-strip sensors produced by Hamamatsu Photonics
\cite{Hamamtsu} have the same geometry of the ones used by
AGILE~\cite{cite:AGILE-Ladder}, but with different thickness, bulk
resistivity and backplane metallization.  The detector size is of $95
\times 95 \times 0.32$ mm$^3$ and each SSD is segmented in 768 strips.
The strips are 48 $\mu$m wide and \mbox{93.196 mm} long with a pitch of
\mbox{121 $\mu$m}. The bulk resistivity is \mbox{$>$ 7 k$\Omega \,
\cdot$cm} with a full depletion voltage of \mbox{55 V} maximum. The
average total leakage current is of \mbox{116 nA} at \mbox{150 V}, well
below the specification of \mbox{900 nA}. The SSDs are glued on the flex
part of the Tracker Front-end Hybrid (TFH) board to form a ladder, as
shown in Fig.~\ref{fig_stk_ladder}. The TFH serves as mechanical support
for the SSDs and for the collection and amplification of the signals
output from the strips. The readout is done one every other strip
(corresponding to 384 channels per ladder), in order to keep a good
performance in terms of spatial resolution, and at the same time reduce
the number of readout channels. The signal shaping and amplification is
performed by six VA140 ASIC chips (produced by IDEAS~\cite{IDEAS}) mounted on the TFH. The chip design is an updated version of the VA64HDR9A chips used in AMS-02~\cite{cite:AMS-VA}. Each VA140 chip reads 64 channels. %

\begin{figure}[t]
\begin{center}
\includegraphics[width=0.7\textwidth,height=0.4\textheight]{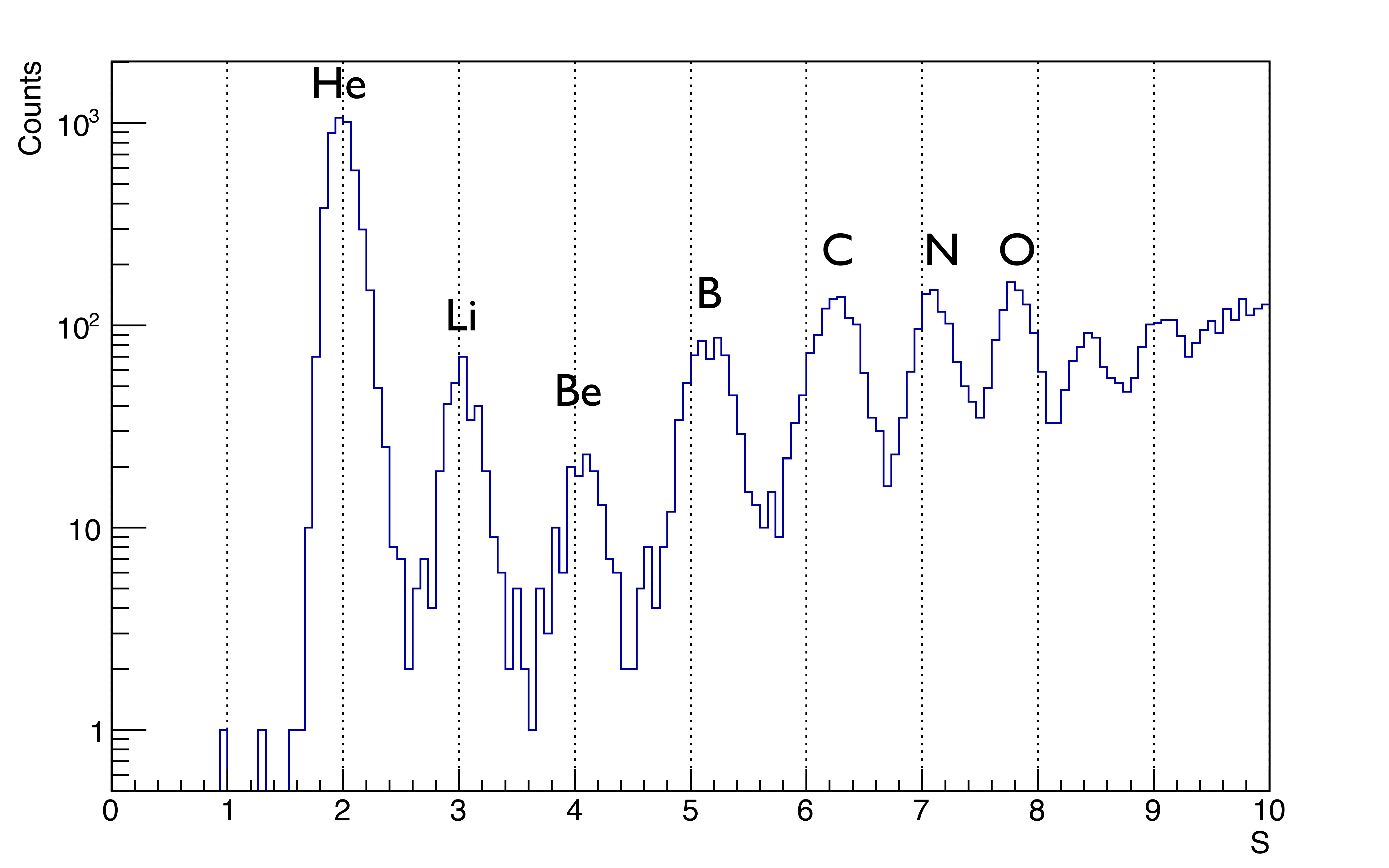}     
\vskip -0.6cm
\caption{STK signal mean distribution for nuclei produced by a lead beam on target, after removing Z=1 particles. The signal mean, with current reconstruction procedures, allows for the identification of ions until Oxygen. A dedicated Z dependent charge calibration is currently being set up (see text). }
\label{fig_stk_charge_number}
\end{center}
\end{figure}

The readout and power supply electronics of the Tracker Readout Boards (TRB)
have been mounted on the sides of the trays as shown in
Fig.~\ref{fig_stk}. Each TRB module reads 24 ladders and is made of
three electronics boards: the power board, the control board, and the ADC
board. The ladders are connected to the ADC board which provides the
conversion of the signal from analog to digital, while the voltage to
the front-end electronics and the silicon bias voltage are supplied by
the power board. The control board is equipped with two field-programmable
gate arrays (FPGAs) which handle not only the communication with the DAMPE
DAQ system, but also the reduction of the data size, thanks to a
zero-suppression and a cluster finding algorithm. More details of the TRB
boards can be found \mbox{in \cite{trb_boards,cite:FPGA}}.

\begin{figure}[h]
\begin{center}
\includegraphics[width=0.6\textwidth]{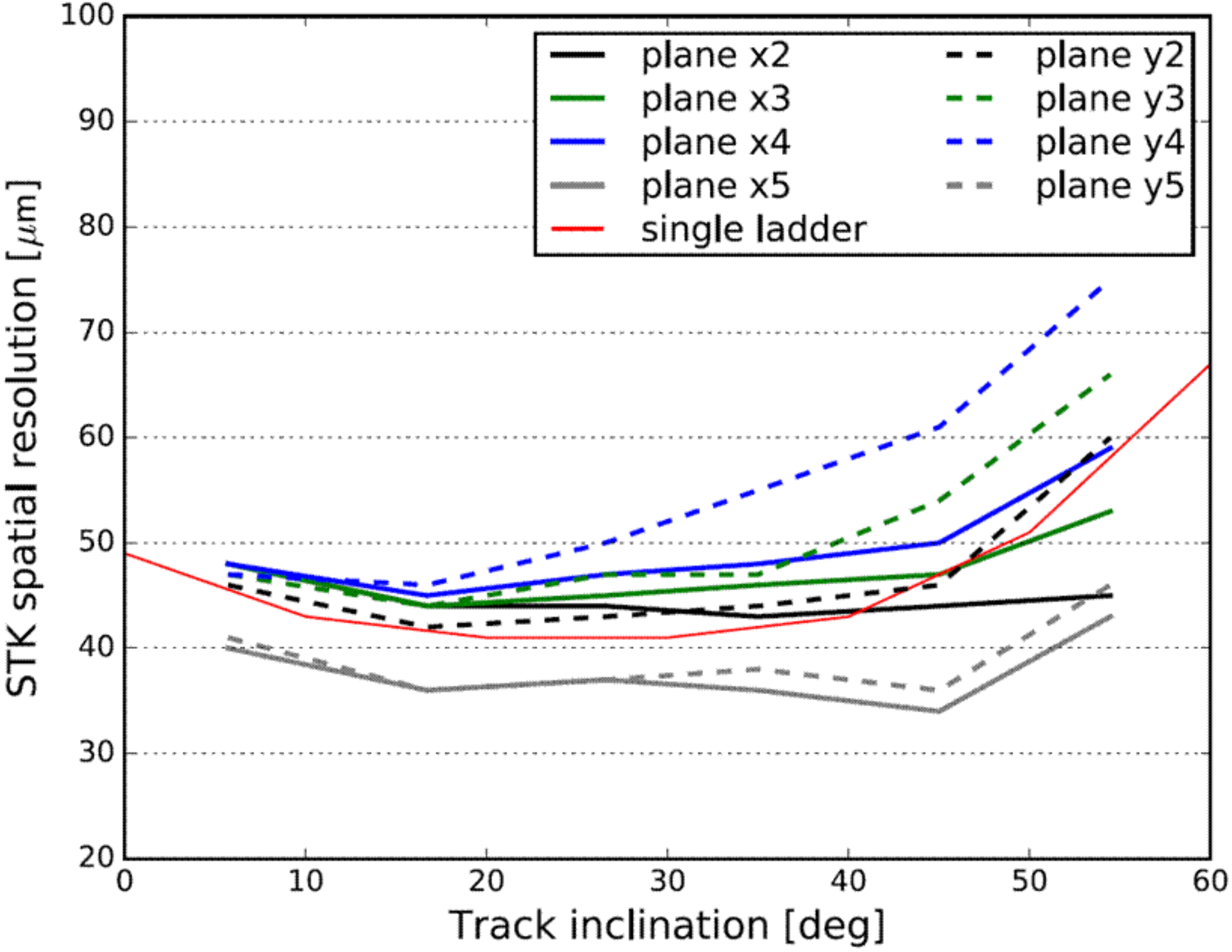}
\caption{Spatial resolution for different STK planes as a function of particle incident angle for cosmic rays data at ground.
The results obtained from a test beam campaign on single ladder are also shown as reference \cite{azzarello_proceeding, cite:Gallo}. }
\label{fig_stk_resolution}
\end{center}
\end{figure}

As discussed in the previous section, several test beam campaigns of the
DAMPE EQM have been conducted at CERN in 2014 and 2015. Moreover, in
order to better characterize the key constituent of the STK, dedicated
tests have been conducted on single ladder units at the CERN Super Proton
Synchrotron facility (SPS). As in the PSD case, the response of the
detector is the same for different singly
charged particles and different energies, as shown in
Fig.~\ref{fig_cluster_charge_response}, while it changes in case of
particles with higher charge numbers (Z $> 1$). The two peaks structure
of the signal distribution, shown on the left side of
Fig.~\ref{fig_cluster_charge_response}, is due to the floating/readout
strip configuration. When a particle crosses a silicon sensor close to a
readout strip and with an incident angle of 0$^\circ$, i.e. orthogonally with respect to the silicon surface, almost all the released charge is collected by a single readout strip (higher charge
peak). On the contrary, when the particle hits a floating strip, only
about 65\% of the original charge is collected by the two nearby readout
strips, which produces the lower charge peak of the ADC distribution. This charge
collection loss reduces as the incidence angle increases, and it could be
recovered with a dedicated correction as function of particle
incident angle and impact position (more details can be found in \cite{azzarello_proceeding, cite:Gallo}). The right panel of
Fig.~\ref{fig_cluster_charge_response} shows the cluster charge
distribution after such correction is applied.

The ions charge identification power of STK was evaluated with a
dedicated test conducted on single ladder units at CERN with a lead
beam. The particle charge can be identified by looking at the mean value of
the signal associated to the track. The signal mean $S={\sqrt{\sum(ADC_i/MIP/N)}}$ is shown in Fig.~\ref{fig_stk_charge_number}. In this formula
$N$ corresponds to the number of clusters composing the track, $ADC_i$
to the signal charge in the $i$-th cluster, and $MIP$ to the cluster
charge in ADC counts of a minimum ionizing particle. This value is
proportional to the particle charge and allows a straightforward
identification of ions up to Oxygen. Due to the non-linearity of the VAs
above a signal of 200 fC, the identification of ions above Oxygen with
the STK becomes non-trivial and on-going work is under preparation to improve the charge identification power. Moreover, in order to equalize the
signal collected by each ladder and to make it independent from the
incidence angle and the particle hit position on the ladder, a
comprehensive and charge dependent STK signal calibration is in progress. Further improvement of the STK charge resolution is expected in the future.

Thanks to a dedicated campaign of extensive cosmic ray data collected on ground,
the STK detector has been aligned before launch, in order to correct
for displacement and rotation of the SSDs with respect to the nominal
position. The alignment procedure will be the subject of a dedicated
paper. Here we only report the spatial resolution as a function of incident angle after alignment, shown in Fig.~\ref{fig_stk_resolution}. As a result of the alignment, the spatial resolution is below 80$\, \mu$m within the angular acceptance of
the STK (i.e. incidence angle $< 60^\circ$) and below 60$\, \mu$m for
particle incidence angles within 40$^\circ$. This result is in agreement with the spatial resolution measurements
obtained in test beam campaigns at CERN SPS on a single ladder~\cite{azzarello_proceeding, cite:Gallo}.

\subsection{The BGO calorimeter (BGO)}

\begin{figure}[!htb]
\centering
\includegraphics[scale=0.4]{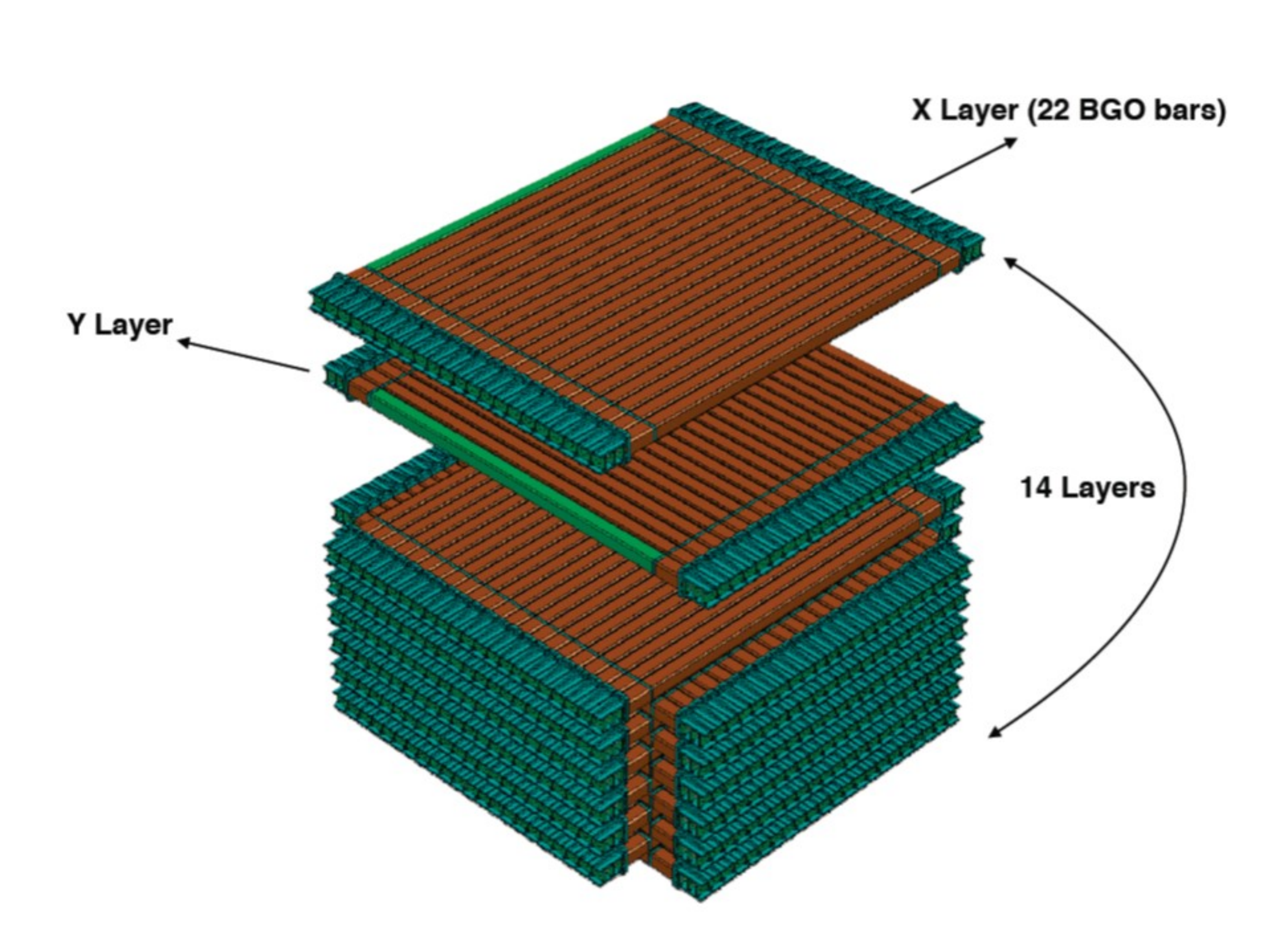}
\caption{Schematic view of the DAMPE BGO calorimeter.} \label{fig:BGOStructure}
\end{figure}

The BGO calorimeter onboard DAMPE has three primary purposes: (1) measuring
the energy deposition of incident particles; (2) imaging the 3D profile
(both longitudinal and transverse) of the shower development, and provide
electron/hadron discrimination; (3) providing the level 0 trigger for
the DAMPE data acquisition system~\cite{Chang2014,Chang2009,WuJ2011,
GuoJH2012,ZhangYL2012,FengC2015,ZhangZY2015,ZhangL2015}.
A summary of the key parameters of the BGO calorimeter is given in
Table~\ref{tab:bgo}. Fig.~\ref{fig:BGOStructure} shows the layout of the
BGO calorimeter.

\begin{figure}[!htb]
\centering
\includegraphics[width=0.70\textwidth,height=0.25\textheight]{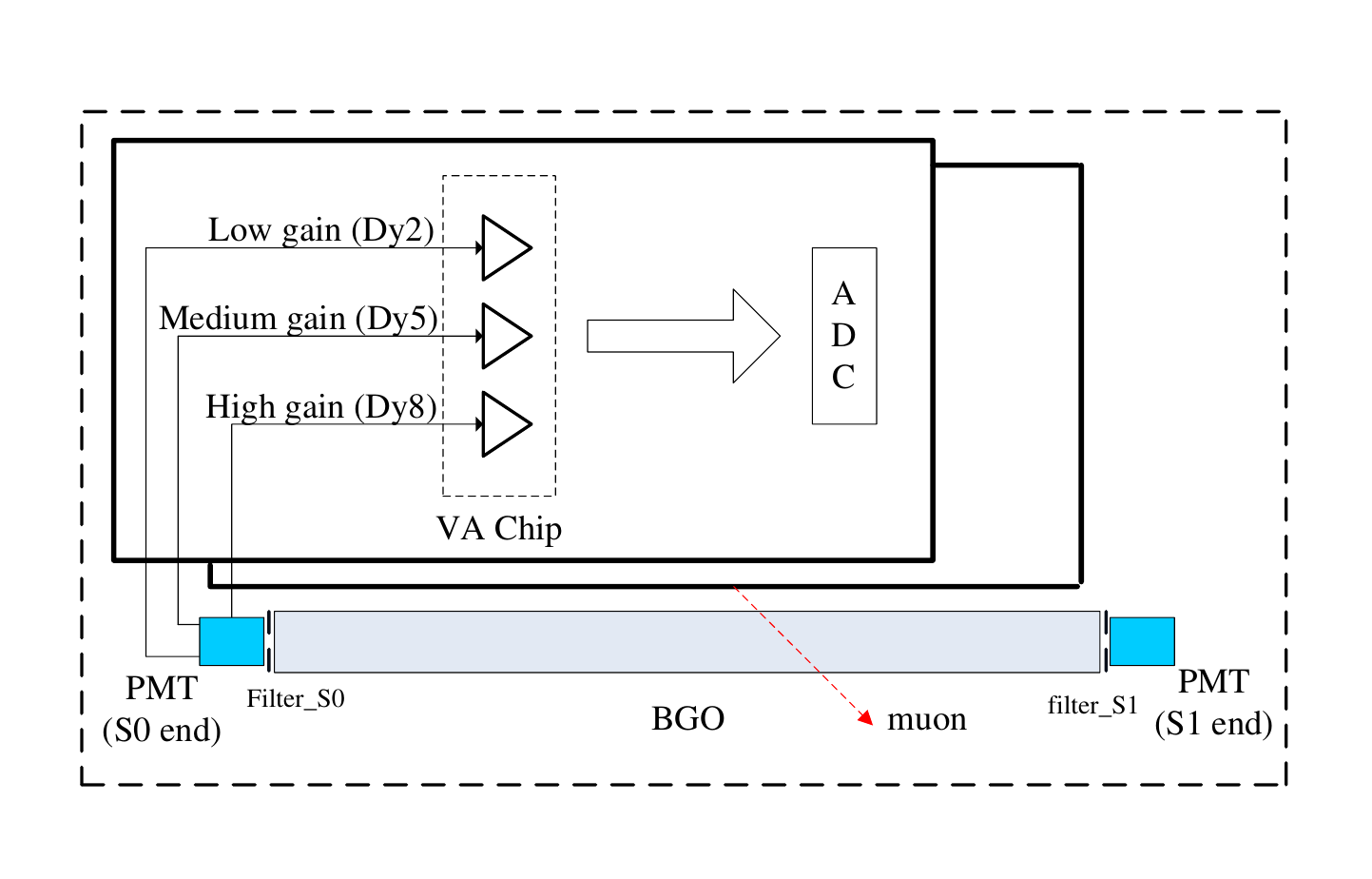}
\caption{The schematic graph of BGO calorimeter readout. The filter on S1 has a 5$\times$ attenuation factor with respect to the one on S0.}
\label{fig:BGOCalReadout}
\end{figure}

\begin{figure}[!htb]
  \centering
   \includegraphics[width=0.48\textwidth,height=0.25\textheight]{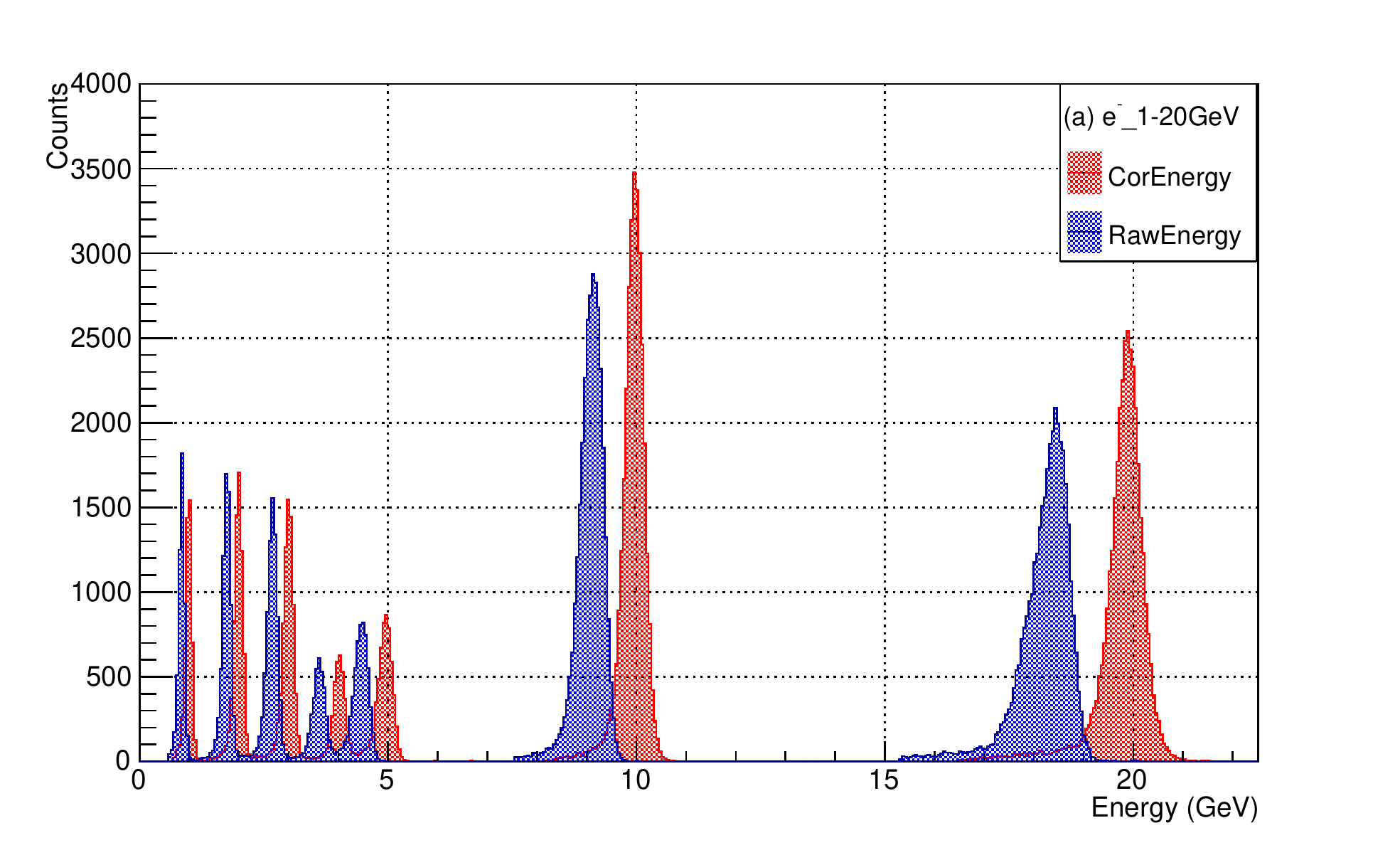}
   \includegraphics[width=0.48\textwidth,height=0.25\textheight]{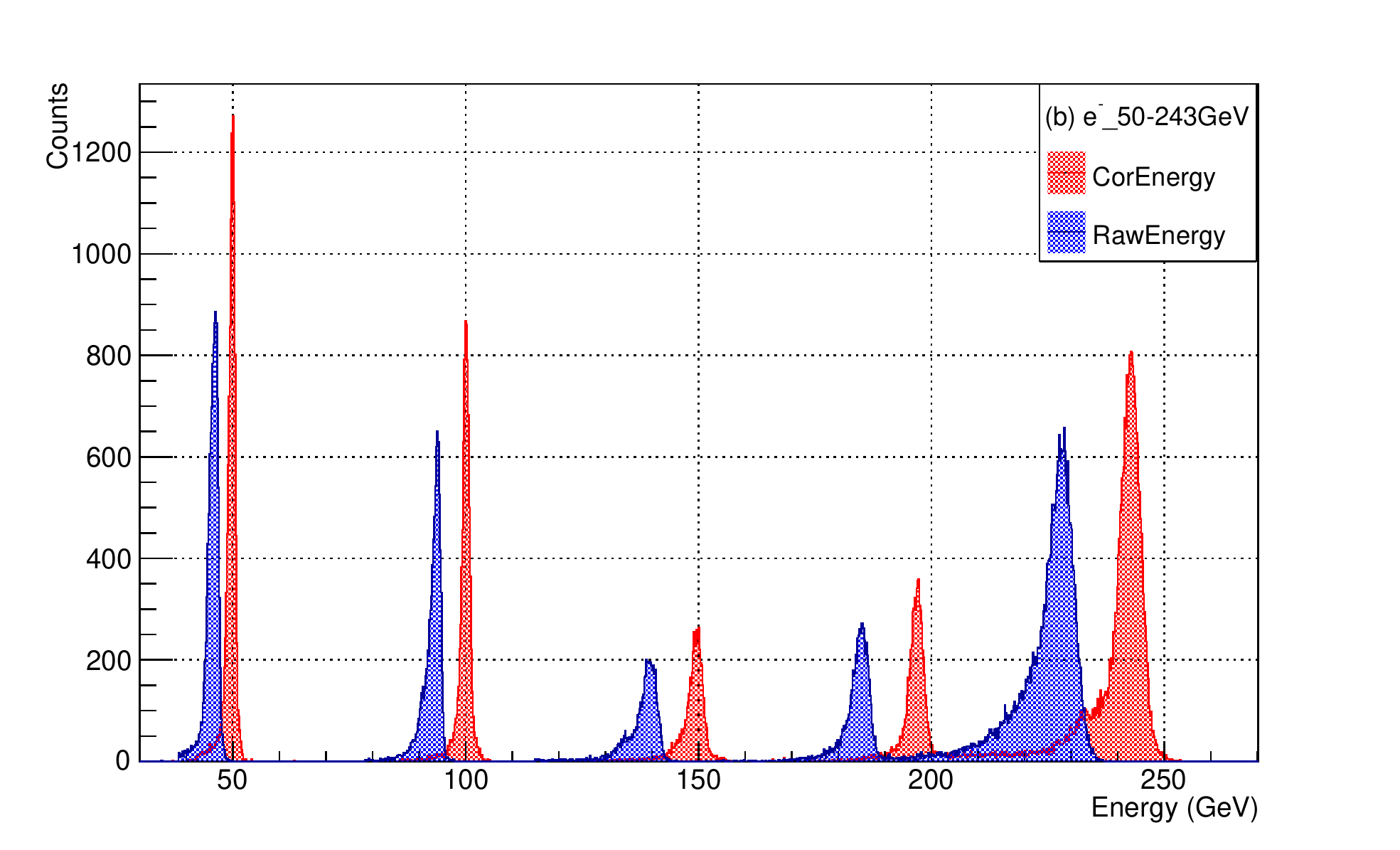}
  \caption{The electron energy distribution from beam test data, before and after correction, as measured in the BGO (see text). the corrected energy resolution is 1.85\% for 10 GeV electrons and 0.80 \% for 100 GeV electrons.}
  \label{fig:energy_beamtest}
\end{figure}

\begin{tiny}
\begin{table}[!ht]
\begin{center}
\caption{Summary of designed parameters and expected performance of the BGO calorimeter.} \label{tab:bgo}
\begin{tabular}{llllll}
\hline
Parameter	                            &	Value                \\
\hline
Active area	     						&   60 cm $\times$ 60 cm (on-axis) \\
Depth (radiation lengths) 			    &   32              \\
Sampling     							& $\geq$ 90$\%$      \\
Longitudinal segmentation               & 14 layers ($\simeq$2.3 rad. lengths each)        \\
Lateral segmentation                	& $\sim$1 Moli\`{e}re radius     \\
\hline
\end{tabular}
\end{center}
\end{table}
\end{tiny}

Each crystal is readout by two Hamamatsu R5610A-01 PMTs (see
Fig.~\ref{fig:BGOCalReadout}), mounted on both ends (named S0 and S1,
respectively). The left/right light asymmetry provides a measurement of
the position of the energy deposit along the bar. The signals are read
out from three different dynodes (dy2, dy5, dy8), thus allowing to cover
a very large dynamic range of energy deposition, The PMTs are coupled
to the crystals with optical filters, which attenuate the scintillation
light produced in the BGO. The filter on S1 has a 5$\times$ attenuation
factor with respect to the one on S0. The high gain readout channels
(dy8) cover the range 2 MeV $-$ 500 MeV (S0 end) and 10 MeV $-$ 2.5 GeV
(S1 end); the medium gain channels (dy5) cover the range 80 MeV $-$ 20
GeV (S0 end) and 400 MeV $-$ 100 GeV (S1 end); the low gain channels
(dy2) cover the range 3.2 GeV $-$ 800 GeV (S0 end) and 16 GeV $-$ 4000
GeV (S1 end). The signals are sent to VA160 chip (or VATA160 for the
bars which generate the trigger see
Sec.~\ref{sec:trig-daq}) which is composed of a charge sensitive
pre-amplifier, a CR-RC shaping amplifier and a sample-hold circuit.
A charge signal can be injected into the front end of the preamplifier
which is used to calibrate and monitor the performance of the VA160.

\begin{figure}
\centering
\includegraphics[width=0.8\textwidth,height=0.35\textheight]{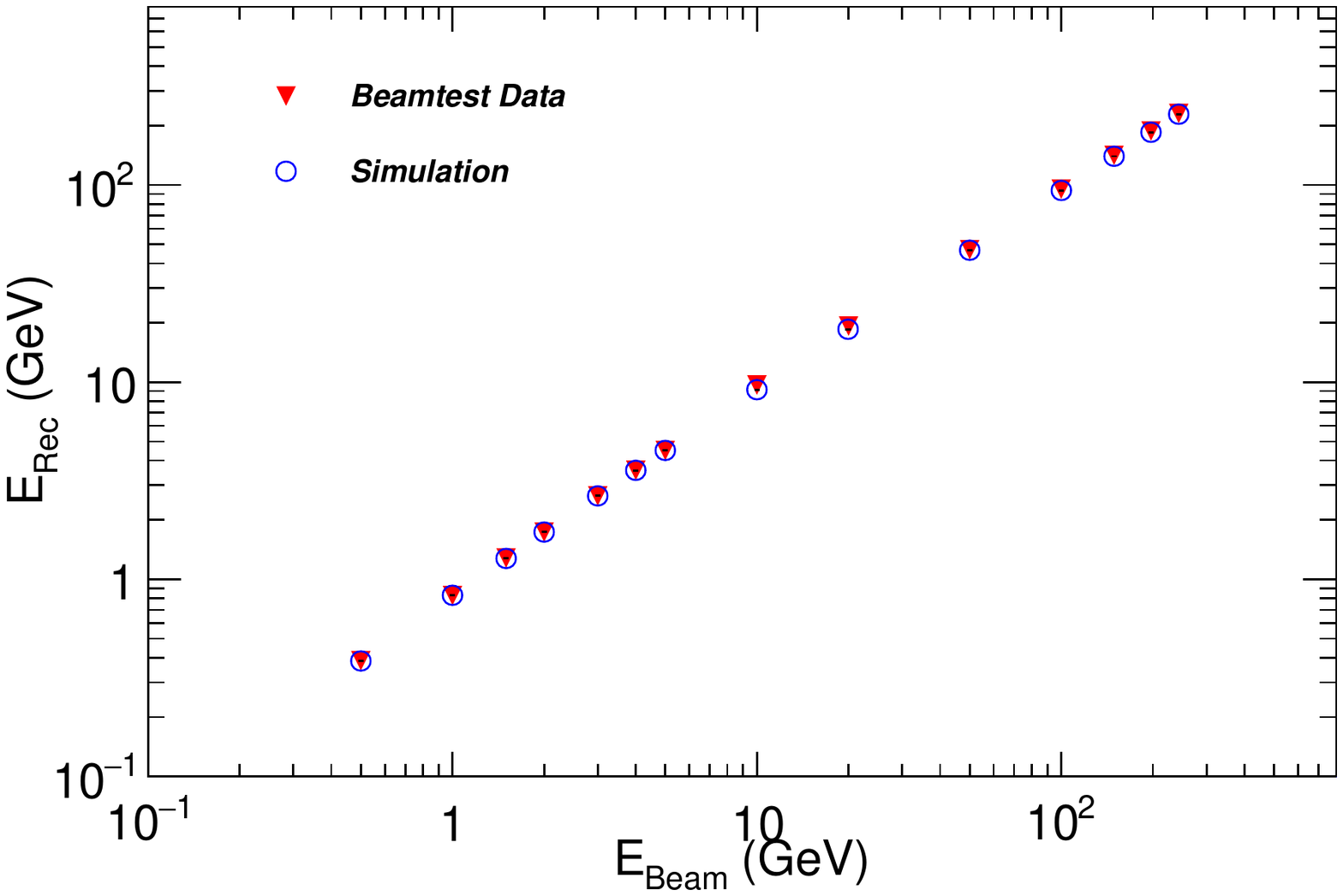}
\caption{Energy reconstructed as a function of the incident energy of electron beam. Red triangles shows the beam test data, and the open blue circles shows the simulation.}
\label{fig:energylinearity}
\end{figure}

The ground calibration of BGO has been performed using both the data collected in a beam test campaign at CERN and cosmic ray data collected from ground.
The calibration procedure includes the measurement of the pedestals, the
evaluation of the calibration constants from the MIP peaks, the evaluation of the dynode ratios, and the measurement of the bar attenuation lengths.
The full details of the calibration procedure are provided in Refs.~\cite{ZhangZY2016,LiZY2016}. Fig.~\ref{fig:energy_beamtest} summarizes the performance of energy reconstruction of the BGO calorimeter for electrons with different energies up to $\sim$ 250 GeV.
The data shown in the figure was obtained during the beam
test campaigns performed at CERN. Details on the energy reconstruction and the electron/proton separation are discussed in Section ~\ref{Sec:EnergyRec} and
Section~\ref{Sec:ElectronProtonSep}.
The linearity of reconstructed energy is better than 1\%, as shown in the Fig.~\ref{fig:energylinearity}.
The energy resolution is better than $1.2\%$ at the energies above 100 GeV (see Fig.~\ref{fig:P-Energyresolution}).

\subsection{The NeUtron Detector (NUD)}

The main purpose of the NUD is to perform electron/hadron identification
using the neutrons produced in hadronic showers initiated in the BGO
calorimeter. In fact, for a given initial particle energy, the neutron
content of a hadronic shower is expected to be one order of magnitude
larger than that of an electromagnetic shower. Once the neutrons are
created, they are quickly thermalized in the BGO calorimeter, and the
total neutron activity over a few microseconds is measured by NUD.
Table~\ref{tab:NUD} summarizes the key parameters of the NUD.

\begin{figure}[!ht]
    \centering
    \includegraphics[scale=0.6]{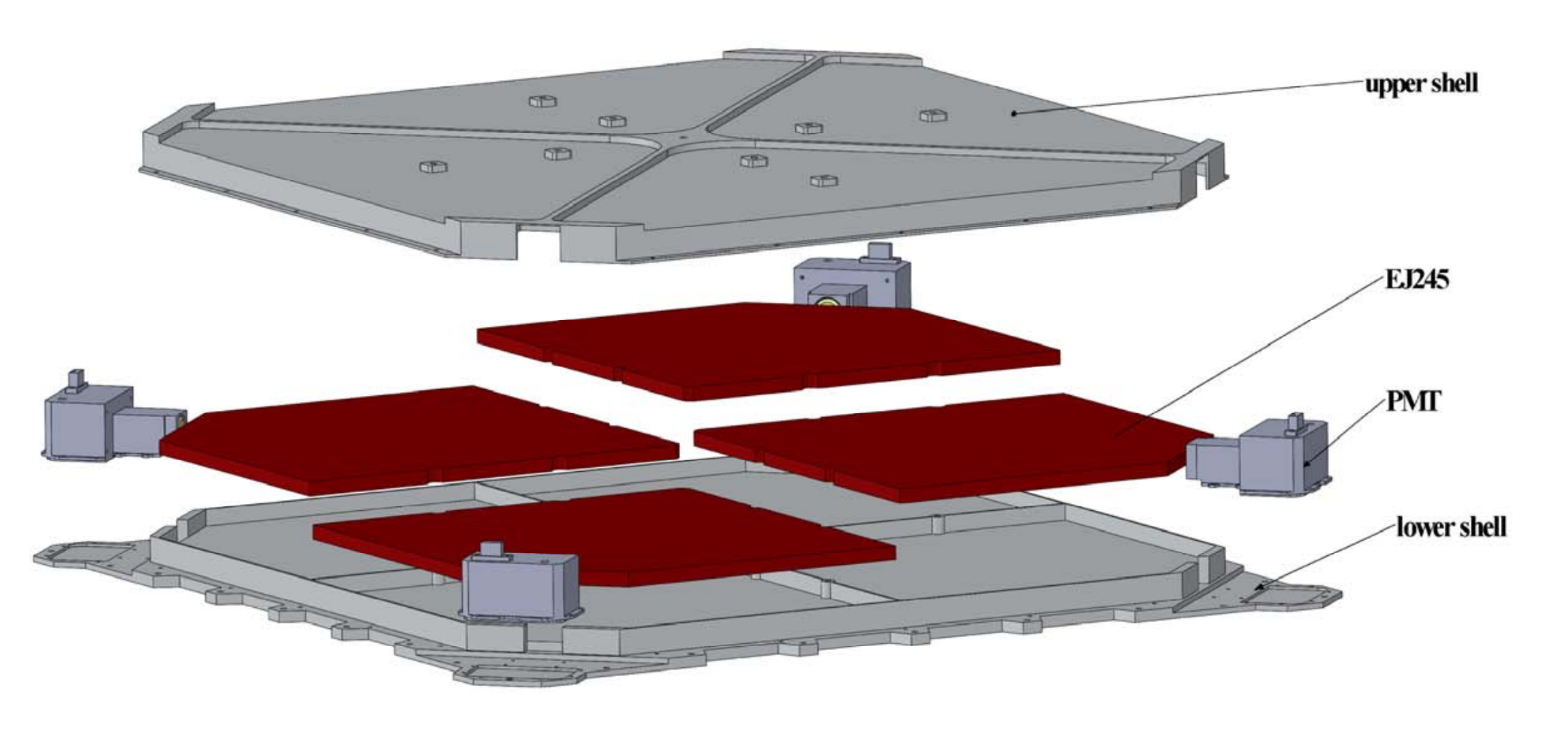}
    \caption{The structure of NeUtron Detector (NUD).} \label{fig:NUD}
\end{figure}

Fig.~\ref{fig:NUD} shows the detailed structure of NUD. It consists of
four $30 {~\rm cm} \times 30{~\rm cm} \times 1.0 {~\rm cm}$ blocks of
boron-loaded plastic scintillator (Eljen Technologies EJ-254), with 5\%
boron concentration by weight which has the natural $^{10}$B abundance of
20\%~\cite{HeM2016}. Each scintillator is wrapped with a layer of
aluminum film for photon reflection, anchored in aluminum alloy
framework by silicone rubber, and readout by a PMT. The space between
plastic scintillators and aluminum alloy framework is 1 mm on each side,
and is filled with silicone rubber to relieve the vibration during the
launch.

The scintillators are embedded with wavelength shift fibers for optical
transmission in order to reduce the fluorescence attenuation and increase
photon collection efficiency, and then the signals are readout by corner-on
Hamamatsu R5610A-01 PMTs. The R5610A-01 is a 0.75 inches diameter head-on,
10-dynode PMT with a maximum gain of $2\times 10^{6}$, and a spectral
response ranging from 300 nm to 650 nm, which is a good match to EJ-254's
425 nm maximum emission wavelength.

Neutron captures are the dominant source of photon generation in the NUD
after $\sim 2\,\mu$s from the initial calorimeter shower. Neutrons
entering the boron-loaded scintillator can in fact undergo the capture
process $^{10}{\rm B}+n\rightarrow ^{7}{\rm Li}+\alpha+\gamma$ with a
probability inversely proportional to their speed, and a time constant
for capture inversely proportional to the $^{10}$B loading. About 600
optical photons are produced in each capture~\cite{Drake1986}.

\begin{figure}[!ht]
    \centering
    \includegraphics[scale=1.2]{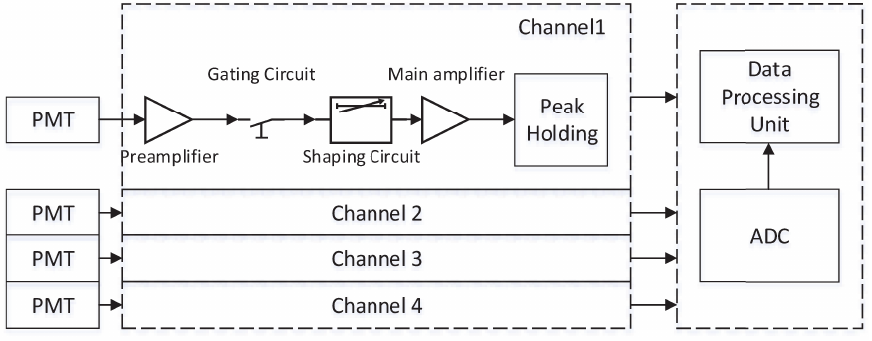}
    \caption{A block diagram of the NUD's Readout Electronics.} \label{fig:NUD-2}
\end{figure}

A block diagram of the readout electronics is shown in Fig.~\ref{fig:NUD-2}.
There are four signal channels provided in one data processing board.
Each channel contains a fast pre-amplifier, a gating circuit (GC),
a shaping circuit (SC) and a main amplifier with peak holding chip (PHC).
The GC and PHC are controlled by the data control unit of the DAMPE
satellite. The GC is designed to prevent any early signal entering the
SC, and is switched-on $1.6~\mu s$ after the triggering signal produced
by BGO. Then the delayed neutron signal could be shaped and amplified to
the PHC. After the ADC finishes the acquisition of all four signals,
a release signal will be sent to the PHC and GC to shut off the signal
channel and wait for the next trigger.

\begin{figure}[!ht]
\centering
\includegraphics[scale=0.35]{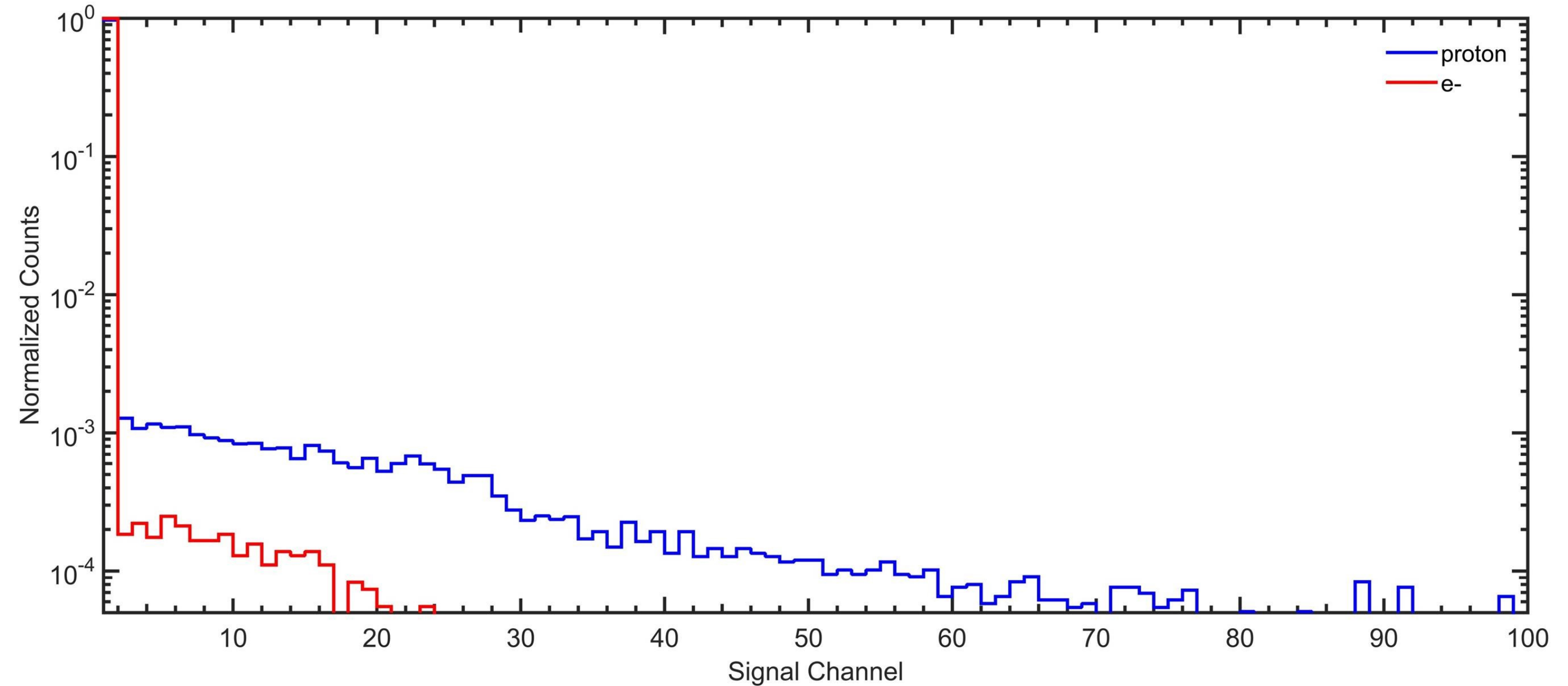}
\caption{NUD signals for protons and electrons with an energy of $\sim 150$ GeV deposited in the BGO calorimeter
(the distributions are normalized to unit area).}
 \label{fig:NUD-4}
\end{figure}

\begin{tiny}
\begin{table}
\begin{center}
\caption{NUD designed parameters.} \label{tab:NUD}
\begin{tabular}{llllll}
\hline
Parameter                                    &    4 Plastic Scintillators ($^{10}$B)                        \\
\hline
Active area                                                         &  61 cm $\times$ 61 cm  \\
Energy range                                             & $2-60$ MeV for single detector                 \\
Energy resolution$^a$                                                    & $\leq$10\% at 30 MeV                        \\
Power                                    & 0.5 W                 \\
Mass                                            & 12 kg                            \\
\hline
\end{tabular}
\end{center}
Note: $^a$$\sigma_E/E$ assuming Gaussian distribution.
\end{table}
\end{tiny}

The electron and proton data collected during the beam test has been
used to study the particle identification power of the NUD. Since
protons deposit in the BGO is about $1/3$ of their initial energy, we
compared 150 GeV electron events with 450 GeV proton events (depositing
$\approx 150$ GeV in the BGO calorimeter). In Fig.~\ref{fig:NUD-4}, the
NUD signals of electrons and protons are compared. The electron signals
are always less than 30 channels, and in most cases are below 2
channels, while the proton signals are remarkably larger.

The PMTs of the NUD and the bottom BGO layer share the same high voltage
module to save electric power and reduce payload weight. As a result,
the NUD works in the high gain mode during on-orbit operation, which
gives a more powerful capability for electron-proton identification.
Detailed GEANT4 simulations suggest a proton rejection power for NUD (in
its full performance) of a factor of $\sim$10, assuming an electron detection
efficiency of 0.95. Preliminary estimates, based on on-orbit calibration
data, show that a rejection power is $\sim12.5$ for incoming particles with
BGO energy deposit above 800 GeV (details will be published elsewhere).

\subsection{Data Acquisition System and Trigger}
\label{sec:trig-daq}
\begin{figure}[!ht]
	\centering
	\includegraphics[scale=0.6]{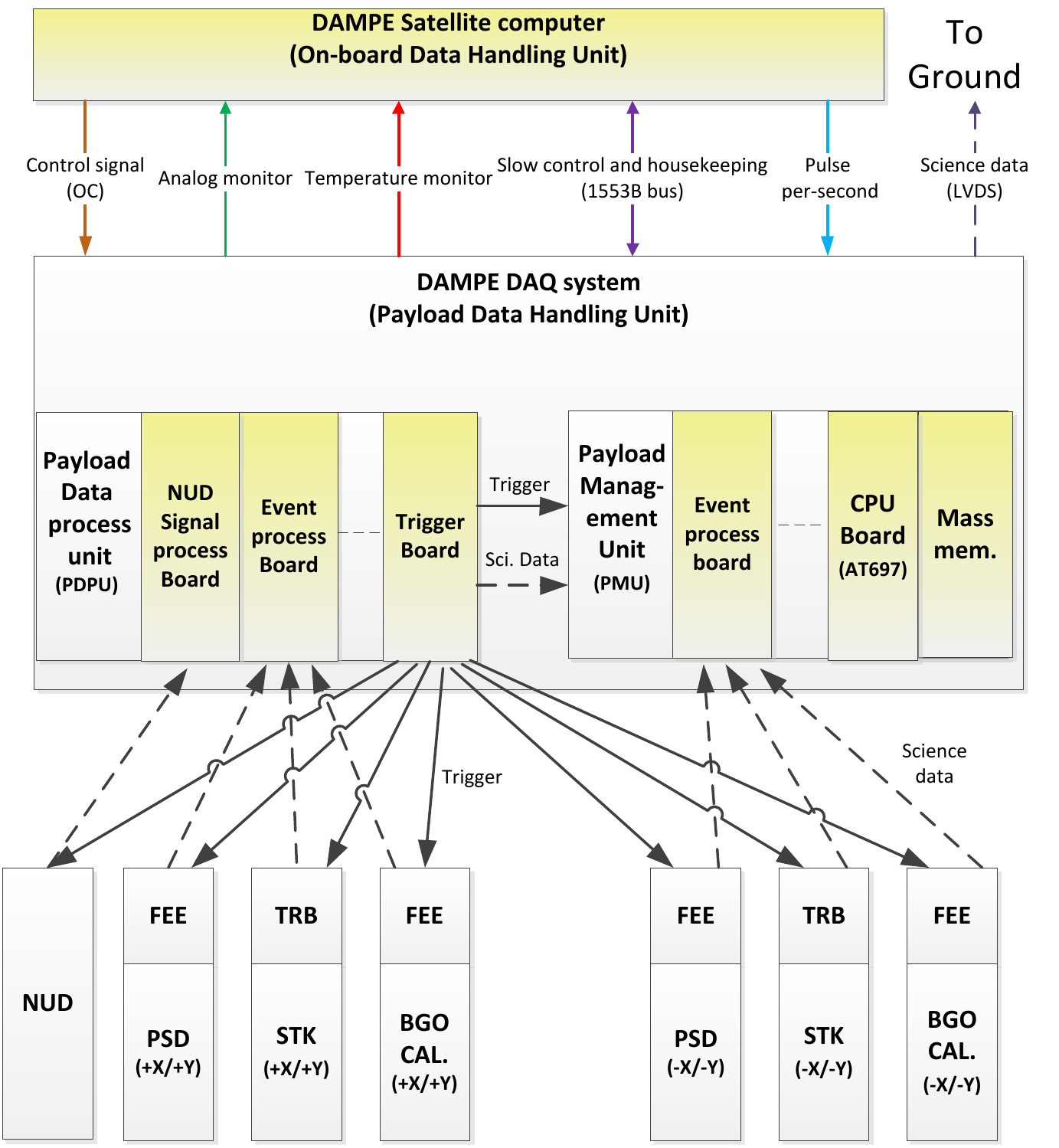}
	\caption{The Data Acquisition System (DAQ) of DAMPE.} \label{fig:DAQ}	
\end{figure}

The data acquisition system (DAQ) receives the commands from the satellite computer, implements trigger decision logic, collects science
and housekeeping data from the detectors, and transfers them to the ground. Fig.~\ref{fig:DAQ} shows the architecture of the DAQ
system. The DAMPE DAQ system \cite{GuoJH2011} consists of two electronics crates, including the Payload Data Process Unit (PDPU) and the Payload Management Unit
(PMU).

The DAQ system is implemented with dual modular redundancy. The PMU is the control center of DAMPE and it is equipped with a 16 GB flash
memory for data storage. The central processing unit (CPU) board of the PMU receives commands from satellite computer through 1553B bus (1 Mbps).
The PMU decodes the commands and distributes them to the PDPU or the FEE of ${\rm -X/-Y}$ sides
directly (see Fig.~\ref{fig:SideView}). When the PMU receives a trigger signal from the trigger board in the PDPU,
it begins to collect science data from the FEE on the ${\rm -X/-Y}$ sides, while data from the FEE on the ${\rm +X/+Y}$ sides
are collected from the PDPU. All collected data are finally stored in the 16 GB mass memory. The PMU also
collects housekeeping data of DAMPE periodically and sends them to the satellite computer. All the science and
housekeeping data are finally relayed to ground with the timestamp of 1$\units{ms}$ precision.
The PMU calibrates its timer with the clock of the Global Positioning System (GPS) spacecrafts with one pulse per second.

The PDPU is responsible for collecting the science data from FEE of ${\rm +X/+Y}$ direction, collecting housekeeping data from FEE, generating global trigger
signal for DAMPE and distributing the commands from PMU.
The trigger board of the PDPU receives signals from the BGO calorimeter and makes a trigger decision within 1 $\mu$s \cite{ZhangL2015}. The
trigger is sent to the FEE and to the PMU, while at the same time the PDPU prevents further events to be collected until all science data is stored, which is collected by the event process board of the PDPU and sent to the PMU.

\begin{figure}[!ht]
	\centering
	\includegraphics[scale=0.9]{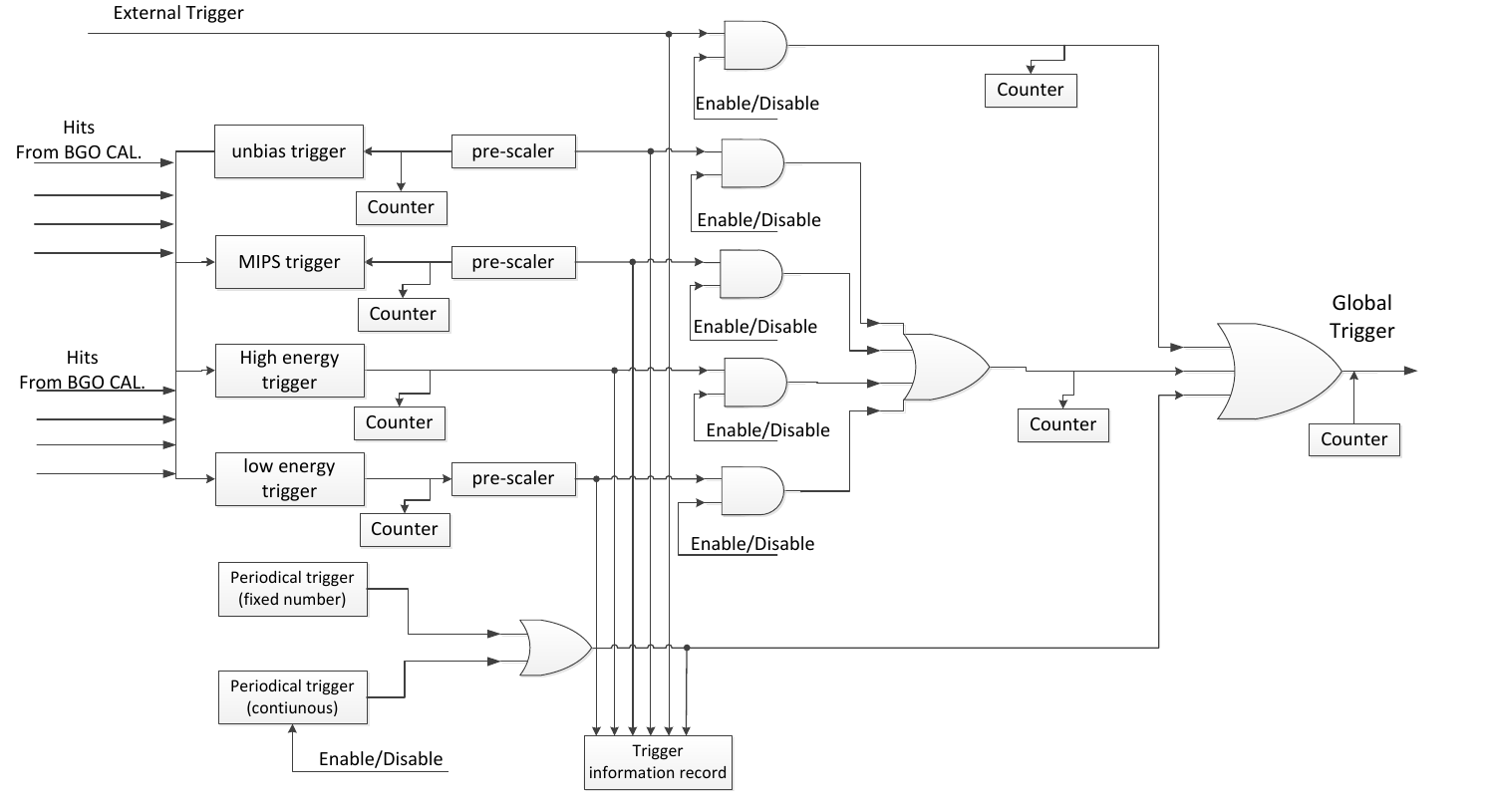}
	\caption{The trigger decision logic of DAMPE (see text).} \label{fig:Trigger}	
\end{figure}

Only the signals from eight out of fourteen BGO layers are sent to the trigger board. The trigger board implements the trigger decision logic with a flash memory based FPGA chip.
Four different triggers have been implemented:
Unbiased trigger, MIP trigger, High Energy trigger and Low Energy trigger.
They are ``OR-ed" to generate the global trigger signal for the detector (see Fig.~\ref{fig:Trigger}).
The Unbiased trigger requires signals in the two top BGO layers exceeding a low threshold of $\sim0.4$MIPs in each hit BGO bar.
The MIP trigger aims to select particles crossing all the BGO layers.
The High Energy trigger selects events with energy depositions in the top four BGO layers exceeding a high threshold of $\sim10$MIPs in each hit BGO Bar.
The Low Energy trigger is similar to the High Energy one, but with a lower threshold of $\sim2$MIPs.
A periodic signal of 100 Hz is also implemented in the trigger board for pedestal calibration.

The Unbiased, MIP and Low Energy triggers are pre-scaled with the ratios of $512:1$, $4:1$, $8:1$, respectively, when the satellite is in the low latitude region ($\pm 20\degrees$). At high latitudes, the MIP trigger is disabled and the pre-scaler ratios of
Unbiased and Low Energy triggers are set to $2048:1$ and $64:1$, respectively.
The expected average rate of global triggers is about $70\units{Hz}$ in flight (the rate of High Energy triggers is $50\units{Hz}$,
the rate of Unbiased triggers is about $2.5\units{Hz}$). The DAQ systems works in an ``event by event'' mode, and a $3\units{ms}$ time interval is set to acquire each event, so that the dead time is fixed to $3\units{ms}$ as a consequence.

\section{Instrument modeling and event reconstruction}\label{sec:modeling}

\subsection{Instrument modeling}

A full Monte Carlo (MC) simulation has been developed to accurately evaluate the detector response to incident particles.
The simulation is central both in the design/optimization phase and in demonstrating the possible achievements
in terms of dynamic ranges, resolutions and background rejection power. The simulation procedure mimics the real
data taking condition of the instrument during both ground tests and in-flight observations, by using
proper input particle fluxes and fully modeling the detector geometry and readout chain.

\begin{figure*}
  \centering
  \includegraphics[scale=0.9]{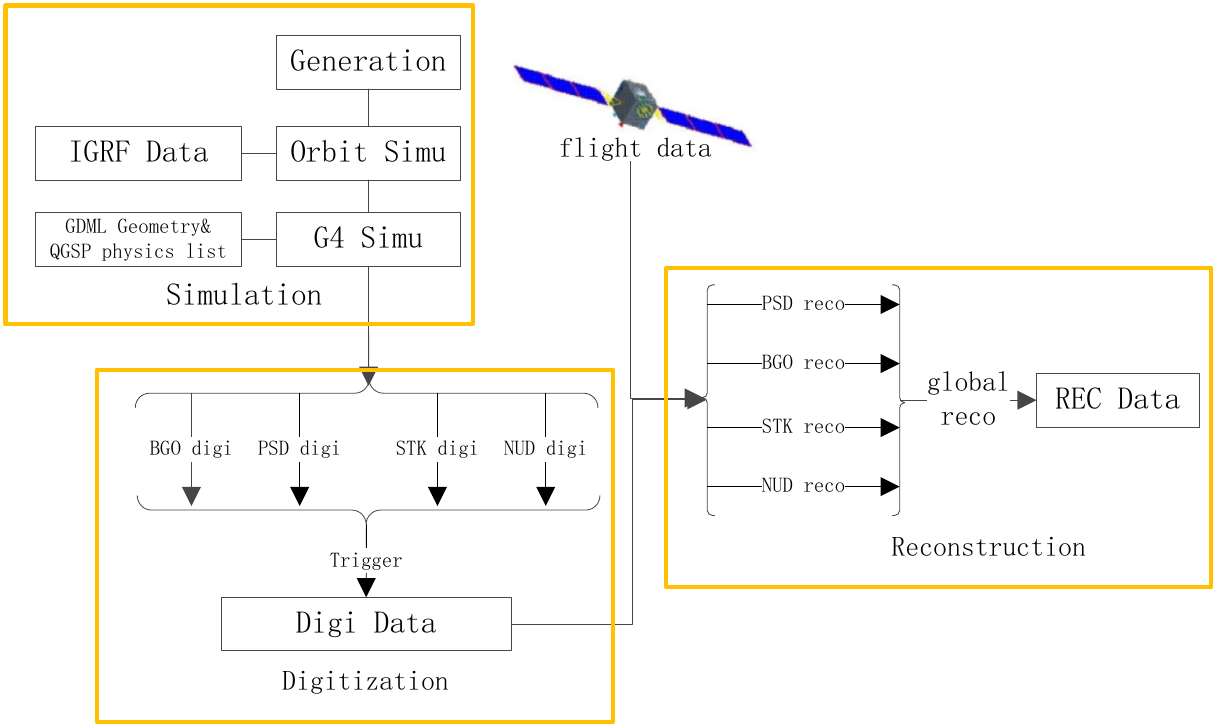}
  \caption{General scheme for DAMPE data handling and simulation.}
  \label{fig:dflow}
\end{figure*}

Fig.~\ref{fig:dflow} shows the flow chart of the data processing for DAMPE, which includes simulation, digitization and reconstruction. The DAMPE simulation
is based on the GEANT4 toolkit~\cite{Agostinelli2003,Allison2006}, a software widely used in high energy physics experiments
to handle particle generation, propagation and interactions. The information on the DAMPE geometry,
including the position and the materials of all the detector elements (both active and passive), is stored in gdml and xml files which are used by GEANT4
to build a detailed model. The whole simulation procedure is implemented in a GAUDI-like software framework~\cite{GAUDI,Tykhonov2015,WangC2016},
which produces collections of energy hits for each sensitive detector element.
A digitization algorithm has been developed to convert energy hits into ADC counts, with the same format as real data,
including the calibration constants (i.e., pedestal noises, PMT gains). In this way, the MC data can be processed by the
same reconstruction algorithms and the simulation can provide an accurate representation of the instrument response for analysis.
Also, for the orbit simulation the same trigger conditions as for real data have been implemented to simulate the final data stream.

\subsection{Event reconstruction}
\subsubsection{Energy reconstruction}\label{Sec:EnergyRec}

The first step of the energy reconstruction algorithm is the conversion
of the ADC counts into energy based on the calibration constants, once
pedestals have been removed, and choosing the signals from the proper
readout dynodes (dy8/dy5/dy2). The total deposited energy is then
calculated by summing up the energies of all BGO crystal elements.
The typical pedestal width is about 8 fC, corresponding to 0.32 MeV (S0)
and 1.6 MeV (S1) for dy8, 12.8 MeV (S0) and 64 MeV (S1) for dy5, 512 MeV
(S0) and 2560 MeV (S1) for dy2, respectively. On orbit, cosmic-ray proton
MIP events will be selected to calibrate the energy response of ADC for
each BGO crystal. The resulting ADC distribution of each individual BGO
crystal will be fitted with a Landau function convolved with a Gaussian
distribution. The most probable value (MPV) corresponds to the MPV in
energy units taken from the simulation ($\approx 21.7\units{MeV}$ for protons).

{\bf Thanks to the multi-dynode readout design, the BGO calorimeter enables a measurement of
 the energy of electrons or gamma rays up to at least 10 TeV without saturation.
The measurable energies for a single bar range from
0.5 MIPs ($\sim11$ MeV) to $10^5$ MIPs ($\sim2$ TeV), covering a dynamic
range of $2\times10^5$. From the simulation we find that, for a 10 TeV
electromagnetic shower, typically the maximum energy deposit in one BGO
bar does not exceed $\sim2$ TeV, which is within the linear region of dy2.}

The energy deposited in the BGO calorimeter underestimates the true energy of incident particles.
Electrons and photons can in fact lose a significant fraction of their energy in the dead materials of calorimeter,
such as the carbon fibers and rubber used for the support structure. For incident electron and photon energies
above hundreds of GeV, the energy leakage should be taken into account.
In addition, the energy deposited in the STK and in the PSD cannot be neglected, in particular for
low-energy incident particles.
The true energy of electrons and photons is evaluated by properly modeling the transversal and longitudinal development
of electromagnetic showers in the calorimeter.

Two methods are used to calculate the corrected energy starting from a set of reconstructed variables, exploiting
their dependence on the deposited energy. In the first case the correction is performed starting from the ratio between
the sum of the maximum energies in each layer and the total deposited energy, which was found to be sensitive to the
energy loss in dead material of BGO calorimeter. In the second case the correction is performed starting from
the depth of the shower maximum obtained by fitting the longitudinal profile with the Gamma-distribution,
which shows a good correlation with the energy leakage.
The correction parameters for different incident energies and different incidence angles are obtained
from the simulations and are checked with beam test data (see Fig.~\ref{fig:energy_beamtest}).
The details of these procedures can be found in ref.~\cite{Yue2016}.

The energy measurements for cosmic-ray protons and nuclei are much more
complicated than that for electrons or gamma rays, as hadronic showers
generally are not fully contained in the BGO. Moreover hadronic showers
include an electromagnetic and a hadronic component with large event-by-event
fluctuations, which brings relatively large uncertainties in the energy
deposition. An unfolding algorithm based on the Bayes
theorem~\cite{D'Agostini1995} will be implemented to estimate the primary
energy spectra of cosmic-ray nuclei.
{\bf DAMPE can measure hadronic cosmic rays to an energy of $\sim100$ TeV
without significant saturation. For such high energy events, the maximum
deposit energy in one BGO bar is typically a few TeV, within the linear
region of dy2. We are developing a correction method using the adjacent
non-saturated bars for a few events which may exceed the linear region of
the readout dynodes.}

\subsubsection{Track reconstruction}

\paragraph{BGO Track reconstruction}
Despite its limited spatial resolution, the BGO calorimeter can also be used for the track reconstruction.
The track reconstruction procedure starts by searching for the ``clusters'' of fired bars in each BGO layer.
A cluster is built starting from the bar with the maximum energy deposit and associating to it all the neighboring
bars on both sides with decreasing energy deposits.
The cluster construction is terminated when one of the following conditions is met: (1) the side
of BGO is reached; (2) a non-fired bar is found; (3) a bar with increasing energy deposit
is found. Finally, we make clusters symmetric about the maximum energy bar.

Therefore, if the left (right) tail of the fired bar cluster has more bars than the right (left) tail,
the bars in excess are removed.
We allow one cluster per layer at most, and then perform a linear fit starting from the positions
of the bars in the clusters, and each bar is weighted with the corresponding energy deposit.
The fitting result, however, is found to bear some systematic bias for inclined incident particles.
To minimize this bias, we rotate the coordinate to align the $X$ axis with the track direction obtained
from the first fit. A second fit is then performed in the new coordinate system, and the final result is obtained by converting back into the original coordinate system. The direction found by the BGO track
reconstruction (if available) is used as a seed for the STK track reconstruction.

\paragraph{STK Track reconstruction}
The raw data of STK are ADC values as the output of data reduction algorithm on board of the satellite~\cite{ZhangF2016}.
Preliminary clusterization of signal is performed on board of the satellite, where cluster seeds are found from the channels which have a
signal-to-noise ratio $S/N>3.5$ and additional strips with $ADC>5$. A refined hit reconstruction is then performed offline from ground, as outlined below.
The ADC values are grouped into arrays of 384 channels per ladder. Channels which did not pass the on-board data reduction are assigned to zero.
The offline clustering algorithm looks for seeds which are defined as local signal
maxima with $S/N>4$, and then form the cluster by collecting all the neighboring strips with $S/N>1.5$.

In order to resolve
multi-peak clusters (which can occur for example in photon conversions into $e^{+}e^{-}$ pairs, where each peak correspond
to its own particle) the cluster reconstruction terminates if a strip signal fulfills the condition $S_{n} / N-S_{n-1} /N>5$,
where $S_{n}$ and $S_{n-1}$ are the signals in the current strip in the cluster and the next strip respectively.
The hits in X and Y projections in same tracking plane are then combined in all possible ways to form three-dimensional
hits. Since quarter-planes of STK are readout by separate electronic boards, only X-Y hit combinations coming from the same
quarter plane are allowed, thus reducing significantly the number of candidate hits.

Track reconstruction is done as follows. The direction found in the BGO is projected onto the closest layer of the STK
with the corresponding error matrix, either infinite, or the one evaluated from the shower position and angular resolution as a function of energy.
If the hit is found within a reasonable window around the projected position, a seed is formed and the track is reconstructed using the Kalman filter.
If the resulting track is of insufficient quality (i.e. the $\chi^2$-test or the number of hits in the track does not fulfill the corresponding threshold values),
the procedure is repeated with other hits in that layer. If a track is not found afterward, it is repeated with the hits in the second and third closest layer
to the calorimeter. If a track is found, the whole procedure is repeated again with the first point of previous track being removed from the list of available points. The same iterations are
repeated from beginning until all seed points are exhausted. Finally, the procedure is repeated also with the three furthermost layers of the tracker in
the opposite direction (towards calorimeter). Once a set of tracks is formed, the ghost tracks are eliminated by looping over all tracks and removing those
with lover quality crossed by the other tracks. The track forks (two tracks starting from the same point) in the direction towards the calorimeter are allowed,
while those which point toward the opposite direction, are considered as a track crossings and treated correspondingly.

\subsubsection{Charge reconstruction}
The measurement of the energy spectra of cosmic-ray nuclei ($Z=1-26$) in the
energy range from $5 \units{GeV}$ to $100 \units{TeV}$ is a major goal of DAMPE.
The charge of cosmic rays can be measured by both the PSD and the STK.

A charged particle crossing a PSD strip loses energy mainly by ionization, with the energy deposition
being proportional to $Z^2$ and to the path length. The first step of charge reconstruction is to find the candidate track,
which allows to find the PSD strips crossed by the particle, and to evaluate the path lengths and the positions in which
the tracks intersect the strips.
Since each PSD strip is readout by two PMTs mounted at each end, two signals per strip are obtained.
From each signal an energy deposition value is calculated, correcting for the path length
and the position of the track along the strip to account for light attenuation.
Since a track can intersect a maximum of four PSD strips, a total of eight energy values per event can be used for charge reconstruction, which are then combined to provide an accurate estimate of $Z$.

The STK, with its 12 layers of silicon strip detectors, can also be used to measure the charge of incident particles, starting from the energy deposition points for the clusters along the track.
The energy deposition for a cluster can be deduced from the impact point and incidence angle.
The impact point can be estimated by the ADC values of the readout strips in the cluster~\cite{Alcaraz2008}.
The charge number can be estimated by combining all those measurements.
Furthermore, in case of fragmentation of an incoming nucleus due to interaction with material of the instrument (for example with the tungsten plates), the charge number is expected to change along the path of the track towards the calorimeter. The PSD and STK will be combined to further improve the measurement of $Z$.

\subsubsection{Electron/proton identification}
\label{Sec:ElectronProtonSep}
The measurement of the total spectrum of cosmic ray electrons/positrons is a major goal of DAMPE. Therefore, besides the track and energy reconstruction, a high identification and discrimination power of protons from electron/positrons is required. The basic approach for electron/proton identification is an image-based pattern recognition method, mainly inherited from the one used in the ATIC experiment~\cite{Chang1999,Schmidt1999,Chang2008a}.

Since the BGO has a radiation length of $1.12\units{cm}$
and a nuclear interaction length of $22.8\units{cm}$, showers initiated by electrons (electromagnetic) and protons (hadronic) will behave very differently in the
BGO calorimeter. Two of the most important features are the radial and longitudinal development of the shower. MC simulation and beam
test data show that electrons and protons can be indeed well separated.
In the GeV-TeV energy range, the proton rejection power can in fact reach a level of $10^5$,
while keeping at least a $90\%$ electron identification efficiency.
Electrons and protons depositing the same amount of energy in the BGO calorimeter can be separated by means of the reconstructed 3D images of the showers.
An electron/proton rejection power close to $2\times 10^3$ while keeping a 94\% electron identification efficiency has been achieved using BGO only beam test data.

For the DAMPE calorimeter, almost all electrons deposit more than $90\%$ of their energy into the calorimeter while protons usually just deposit $\sim 1/3$.
Since the cosmic ray proton spectrum is approximately proportional E$^{-2.7}$, the on-orbit rejection power will be improved of a factor $\approx 7$
(i.e., $3^{1.7}\approx 7$).
In addition, the High Energy trigger (see Sec.\ref{sec:trig-daq}) has been optimized to suppress the proton events by a factor of $\sim 3$. Finally, the NUD can be used to further increase the rejection power by a factor of $\sim$2.5 at TeV energies. As an independent check, we also adopt the Toolkit for Multivariate Data Analysis (TMVA) and deep learning techniques to perform the e/p identification, which give rather similar rejection powers.

\section{Performance and Operation}

\subsection{Expected performance and tests}
\begin{figure}[!ht]
	\centering
	\includegraphics[scale=0.5]{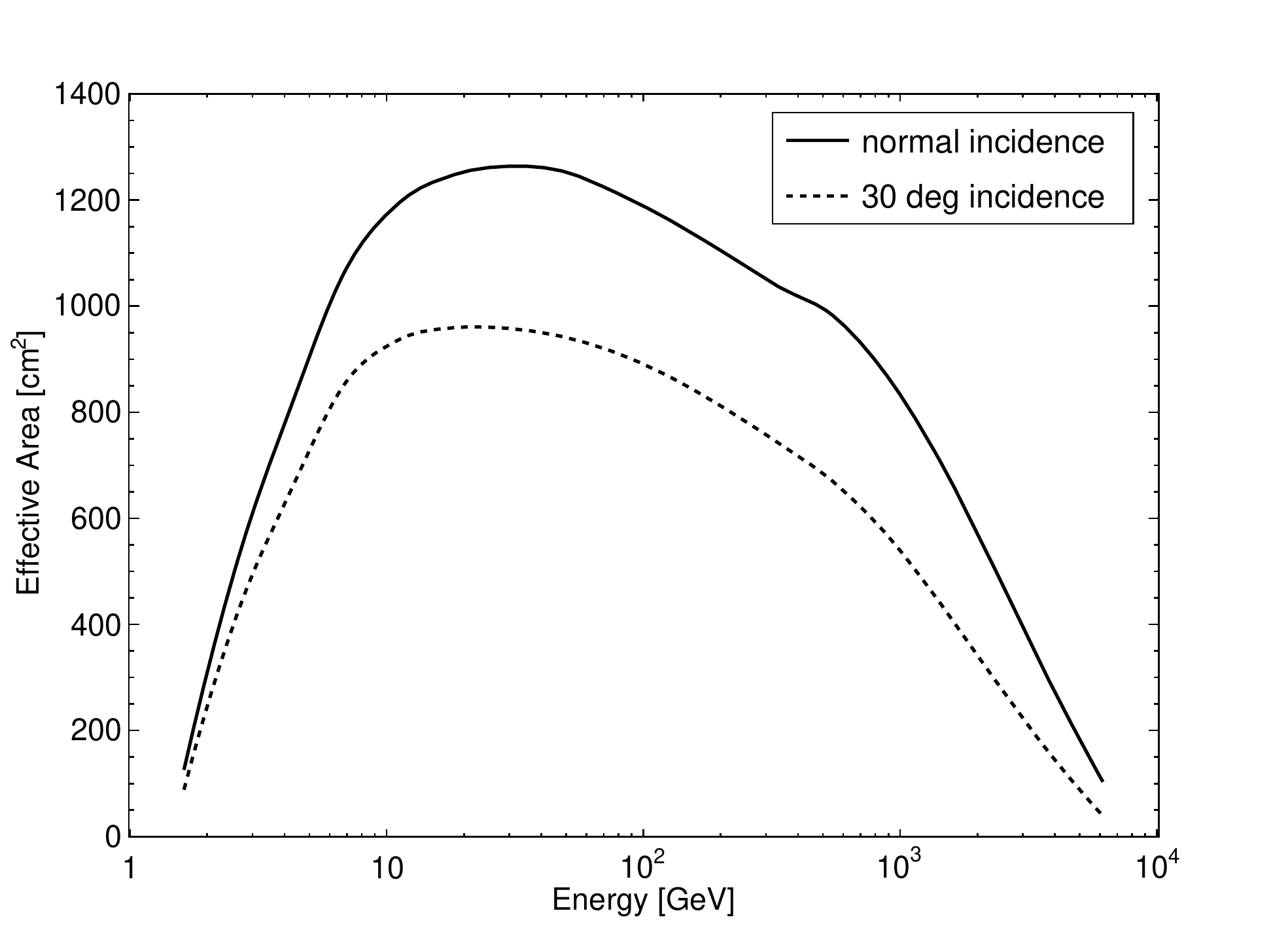}
	\caption{Effective area as a function of energy for gamma rays at normal incidence (solid
line) and at $30^{\circ}$ off-axis angle (dashed line).} \label{fig:P-EffectiveArea-gamma}	
\end{figure}

\begin{figure}[!ht]
	\centering
	\includegraphics[scale=0.5]{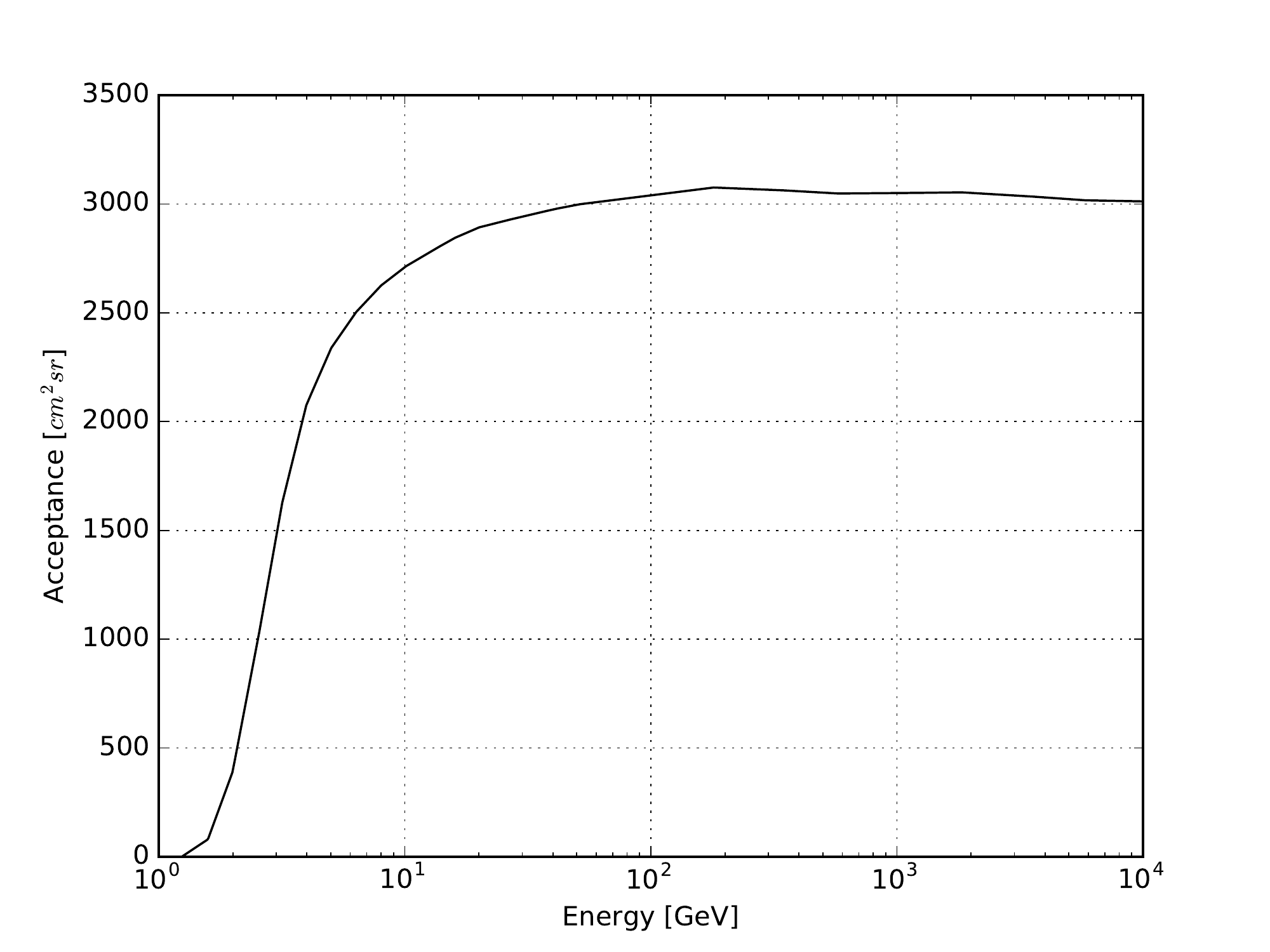}
	\caption{Acceptance for electrons/positrons as a function of energy.} \label{fig:Acceptance}
\end{figure}

\begin{figure}[!ht]
	\centering
	\includegraphics[scale=0.5]{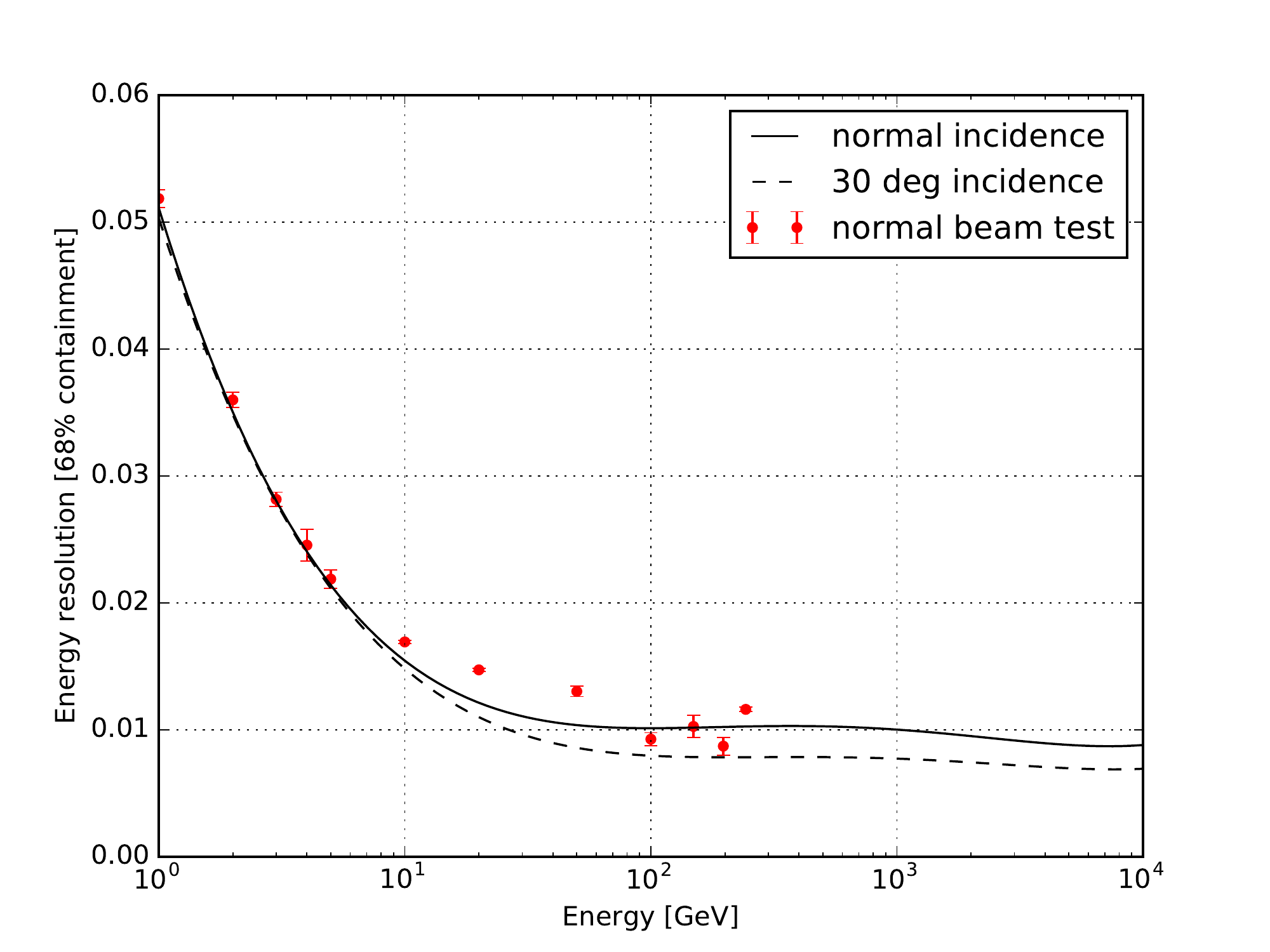}
	\caption{Energy resolution for gamma rays and electrons/positrons at normal incidence (solid line) and
at $30^{\circ}$ off-axis angle (dashed line). DAMPE beam test results (with electrons) are over-plotted as reported in Fig.~\ref{fig:energy_beamtest}.}
\label{fig:P-Energyresolution}
\end{figure}

\begin{figure}[!ht]
	\centering
	\includegraphics[scale=0.38]{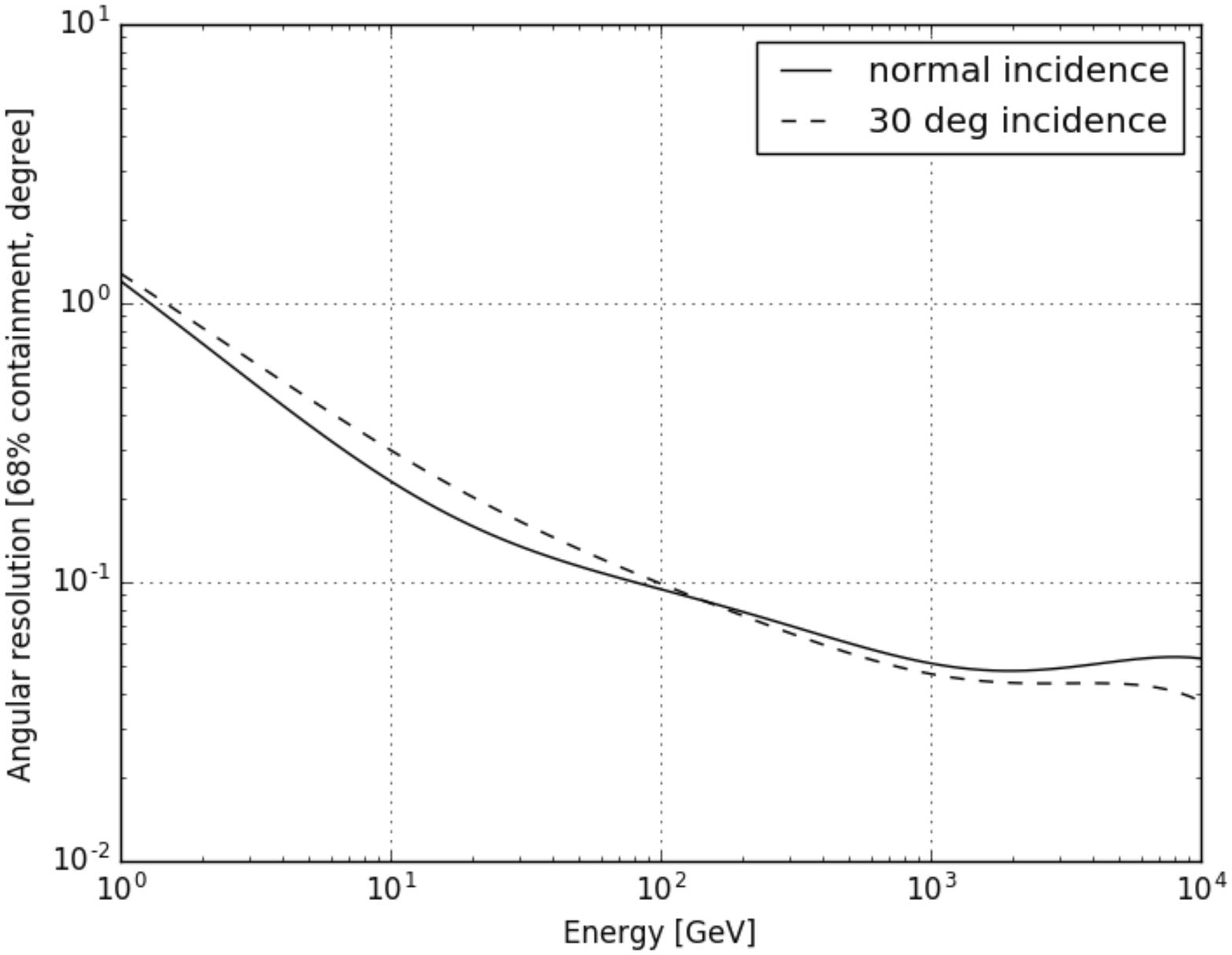}
	\caption{Angular resolution at 68\% containment angle for gamma rays at normal incidence (solid curve) and at $30^{\circ}$ off-axis angle (dashed curve).}
\label{P-AngularResolution}	
\end{figure}

\begin{figure}[!ht]
	\centering
	\includegraphics[scale=0.6]{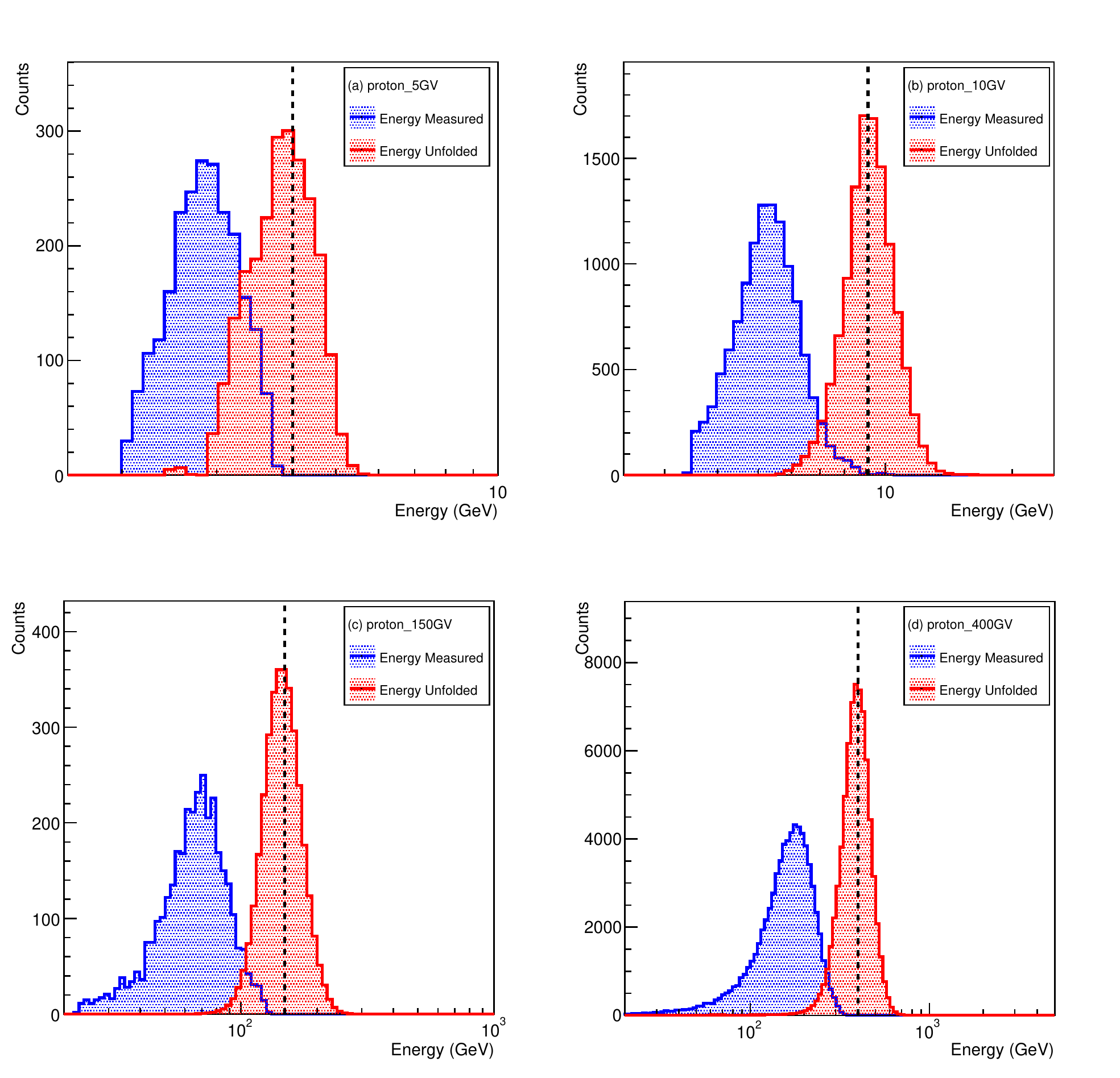}
	\caption{Distributions of deposited energies (blue) and unfolded ones (red) for beam test protons at incident momenta of 5, 10, 150, and 400 GeV/$c$.}
\label{BeamTest_Proton}	
\end{figure}

\begin{figure}[!ht]
	\centering
	\includegraphics[scale=0.5]{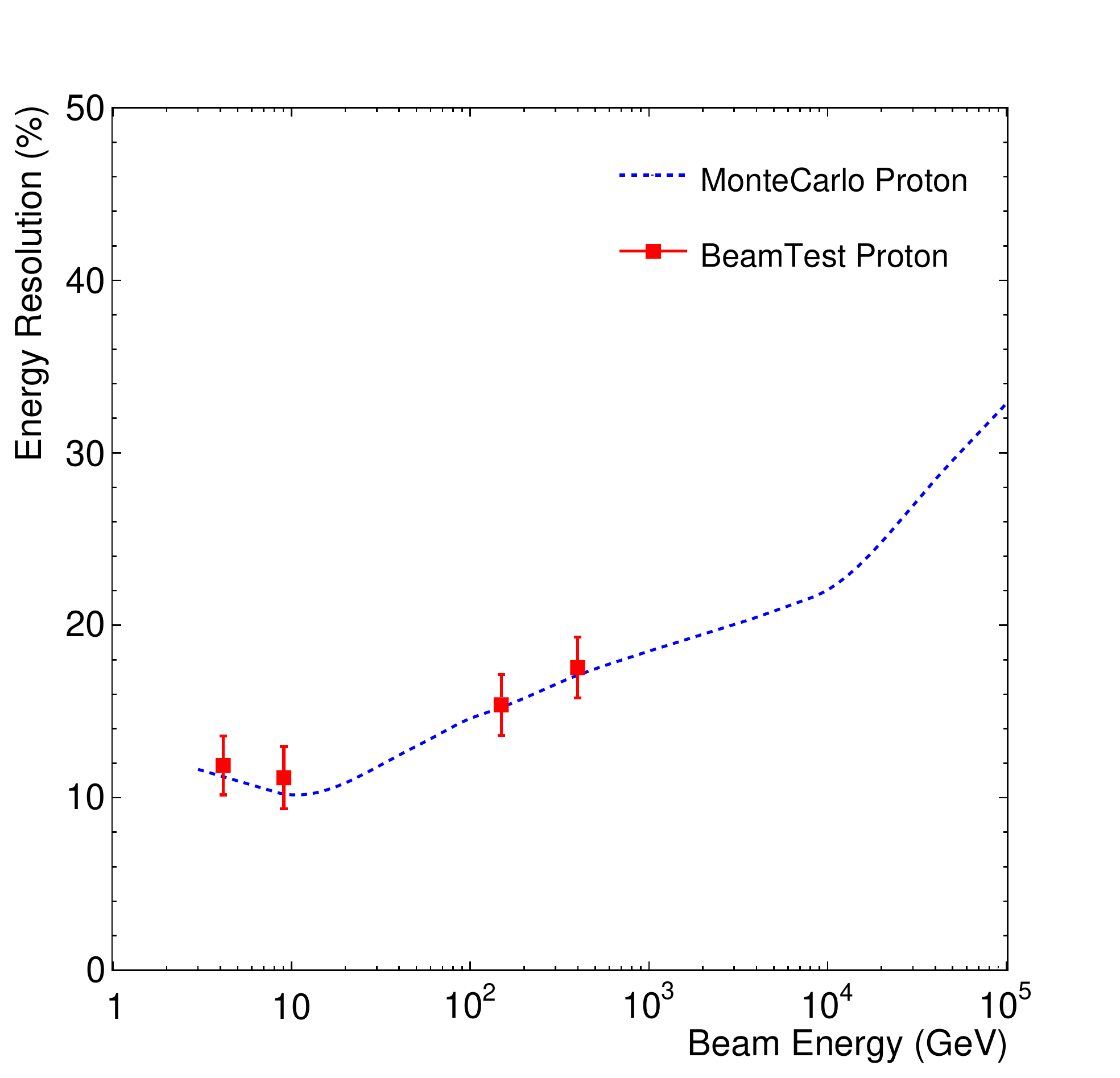}
	\caption{The energy resolution for on-axis protons.  The dotted line represents the energy resolution of MC simulated protons after spectral unfolding while the red points represent the beam test data.}
\label{EneRes_Proton}	
\end{figure}

The expected instrument performance is summarized in Figs.~\ref{fig:P-EffectiveArea-gamma}$-$\ref{P-AngularResolution} for electrons/photons, and Figs.~\ref{BeamTest_Proton}-\ref{EneRes_Proton} for protons. These results are based on simulations of DAMPE instrument performance from the event reconstruction and selection algorithms, which includes trigger filter, track reconstruction, geometry constraints, charge reconstruction, particle identification and energy reconstruction.
The efficiency of each step has been carefully studied with MC simulations and checked with beam test data and cosmic-ray muon data at ground.
The performance parameters (in particular for gamma ray detection efficiency) are expected to improve in the future with improved algorithms, as the event reconstruction and selection algorithms will be further optimized after a better understanding of the on-orbit performance.

Fig.~\ref{fig:P-EffectiveArea-gamma} shows the effective area as a function of energy for gamma ray detection at normal incidence and at $30^{\circ}$ off-axis angle, respectively. The adopted event selection algorithm for gamma rays is the following.
Firstly events with shower well contained in the calorimeter are selected, then a first hadronic background rejection is performed by using information from the BGO only (see Sec.~\ref{Sec:ElectronProtonSep}). Candidate electron/gamma-ray events with a track in the STK are then selected. Finally the PSD is used as an anti-coincidence detector to reject charged particle events. The drop of effective area above $100\units{GeV}$ (shown in Fig.~\ref{fig:P-EffectiveArea-gamma}) is due to the backsplash effect, which has not been taken into account in the present gamma-rays event selection. The same cuts without the PSD anti-coincidence veto can be used to select electrons/positrons. Starting from events with the High Energy trigger, the resulting acceptance for electrons is larger than $0.3\units{m^2~sr}$ above $50\units{GeV}$, as shown in Fig.~\ref{fig:Acceptance}.
The energy resolution for electromagnetic showers is shown in Fig.~\ref{fig:P-Energyresolution}. The angular resolution (i.e. the corresponding 68\% containment angle) for gamma rays converted in the STK is shown in Fig.~\ref{P-AngularResolution} for normal and 30$^{\circ}$ incidence angles, respectively.

{\bf For hadronic cosmic rays, the acceptance is about $0.1\units{m^2~sr}$
for energies above $\sim100\units{GeV}$, which varies for different nuclei
species due to different trigger efficiency. The energy measurement of
cosmic ray nuclei is more complicated than that of electrons/photons,
because of the energy leakage due to limited nuclear interaction thickness
of the calorimeter ($\sim1.6$ nuclear interaction length) and fluctuation
of the hardonic shower development. To convert the measured energy spectrum
to the primary energy spectrum, an unfolding algorithm could be used to
reconstruct the nucleus energy spectrum, by using the MC detector response
matrix. Figure~\ref{BeamTest_Proton} shows the deposit (blue) and
reconstructed (red) energy distributions for on-axis incident proton beams
with momenta of 5, 10, 150, and 400 ${\rm GeV}/c$. The reconstructed procedure allows to recovery the incident beam energy as well.

The energy resolution ($\sigma_E/E$) of on-axis incident protons (after
the unfolding), estimated from the simulation data, is shown by the dotted
line in Fig.~\ref{EneRes_Proton}. As a comparison, the results for the
beam test data at four energies are overplotted. It is shown that the
energy resolution for protons varies from $\sim10\%$ at several GeV to
$\sim30\%$ at $100$ TeV.
Above 10 TeV, the uncertainties on the hadronic interaction model as implemented in Geant4 are expected to be non-negligible. While a detailed treatment is currently being undertaken in the collaboration, we expect these uncertainties to yield uncertainties in the reconstructed spectrum of about 10\%. We are also investigating the use of alternative simulation packages that incorporate hadronic interactions at the TeV scale better (e.g. Fluka \footnote{http://www.fluka.org/fluka.php}).
}

As discussed above, the verification of the estimated performance was carried out using the data from the beam test campaign, as well as a set of data collected with cosmic-ray muons at sea level. In particular, several cosmic-ray muon tests were performed during different stages of the DAMPE assembly, especially in the environmental testing phase and in the pre-launch preparation of the satellite. In these tests, a proper trigger logic was adopted to select cosmic-ray muons.
We were able to collect a large amount of muon events, which has been used to perform a full calibration of the energy response for MIPs and to implement the alignment procedure for the STK. After launch, the spacecraft entered the {\it sky-survey} mode immediately, and a dedicated calibration of the detector was performed in the first 15 days, including pedestals, MIP responses (protons), alignments, and timing etc. Comparison between on-orbit data with simulations and ground cosmic-ray data demonstrates the excellent working condition of DAMPE detectors. {\bf Details of the on-orbit calibration and performance evaluation will be published elsewhere \cite{On-Orbit-Cal2017b}.}

\subsection{Operation}
Since December 17$^{\rm th}$ 2015, DAMPE is orbiting in solar synchronous mode, with each orbit lasting 95 minutes.
The trigger configuration and the pre-scaling factors for the on-orbit science operation have been illustrated in Sect.~\ref{sec:trig-daq}, and ensure a global trigger rate around $70\units{Hz}$. The pedestal calibration is performed twice per-orbit, and all data are regularly transmitted to ground.

On ground the data are processed by the Ground Support System (GSS) and the Scientific Application System (SAS).
Binary raw data (housekeeping and science data) transmitted to ground are first received by three ground stations located in the south, west and north of China
at early morning and afternoon of each day respectively, when the satellite passes China's borderline. Then all binary data are automatically transmitted to the GSS
located in Beijing, and are tagged as level-0 data. On average, about 12 GB level-0 data are produced per day. Upon arrival of the level-0 data at the GSS, they are immediately processed and several operations are performed, including data merging, overlap skipping and cyclic redundancy check (CRC) which is an error-detecting code based on the protocol CRC-16/CCITT.

The level-0 data are daily processed into level-1 data, which includes 13 kinds of completed telemetry source packages, one for science data and 12 for housekeeping data.  Daily level-1 data will then be processed by the GSS within 1 hour. The SAS located at the Purple Mountain Observatory of Chinese Academy of Sciences in Nanjing monitors the level-1 data production 24 hours a day continuously. The new level-1 data will be synchronized to the mass storage at the Purple Mountain Observatory immediately. Then 12 housekeeping data packages
are parsed and inserted into the housekeeping database, which allows to monitor the conditions of the DAMPE payload and the satellite platform. After processing the housekeeping data, routine checks on key engineering parameters are performed to guarantee the proper data taking conditions.

The processing pipeline of science data includes the Raw Data Conversion, Pre-Reconstruction and Reconstruction algorithms
implemented in the DAMPE software framework (DAMPESW). The Raw Data Conversion algorithm splits raw science data into about 30
calibration files and 30 observation files, and converts them into ROOT data files~\cite{ROOT}. During this procedure,
key housekeeping data required by science analysis are also stored into in the ROOT data files. Calibration files are used to extract calibration constants which are used in Pre-Reconstruction and Reconstruction algorithms. Reconstructed data from all sub-detectors are then merged to generate level-2 science data products. These two procedures increases the raw science data volume by approximately a factor of five.

The processing pipeline of science data is designed to run on a cluster of batch processors. The SAS hosts more than 1400 computing cores at the Purple Mountain Observatory, which can reprocess three years of DAMPE data within two weeks. In addition, INFN and University of Geneva computing resources are also used, which are mainly dedicated to MC data production and could also be used as backup reprocessing sites if needed.

\section{Key scientific objectives}
\label{sec:KSO}

DAMPE is a high energy cosmic-ray and gamma-ray observatory with a broad range of scientific objectives. The data sets provided by DAMPE could be used to study cosmic-ray physics, to probe the nature of dark matter, and to reveal the nature of high energy gamma-ray phenomena. The large field of view of DAMPE provides the opportunity to monitor the violent GeV-TeV transients for various purposes.

\subsection{Understanding the acceleration, propagation and radiation of cosmic rays}
\label{sec:CRProp}
Cosmic rays impinging the Earth with energies below $\sim10^{17}\units{eV}$
are believed to be mainly produced through energetic astrophysical processes within the Milky Way. Their interactions with interstellar medium, interstellar radiation fields, and Galactic magnetic fields are the main source of the detected Galactic diffuse gamma-ray emissions. Moreover, cosmic rays are the only sample of matter originated from distant regions of the Galaxy that can be directly measured with spaceborne experiments. Therefore, understanding the origin, acceleration, and propagation of cosmic rays is a crucial subject on the understanding of the Universe.

With more than three years of operation, DAMPE will be able to
observe electrons/positrons or photons from GeV to  10TeV,
and protons, helium or heavier nuclei from 10 GeV to 100 TeV.
The measurement of energy spectra with unprecedented precision and energy coverage at higher energies, together with spatial distribution of these particles are expected to significantly enhance
our understanding of the origin of cosmic rays. Below we outline the key scientific outputs regarding cosmic-ray studies potentially achievable with DAMPE.

\begin{figure*}[!htb]
\centering
\includegraphics[width=0.49\textwidth]{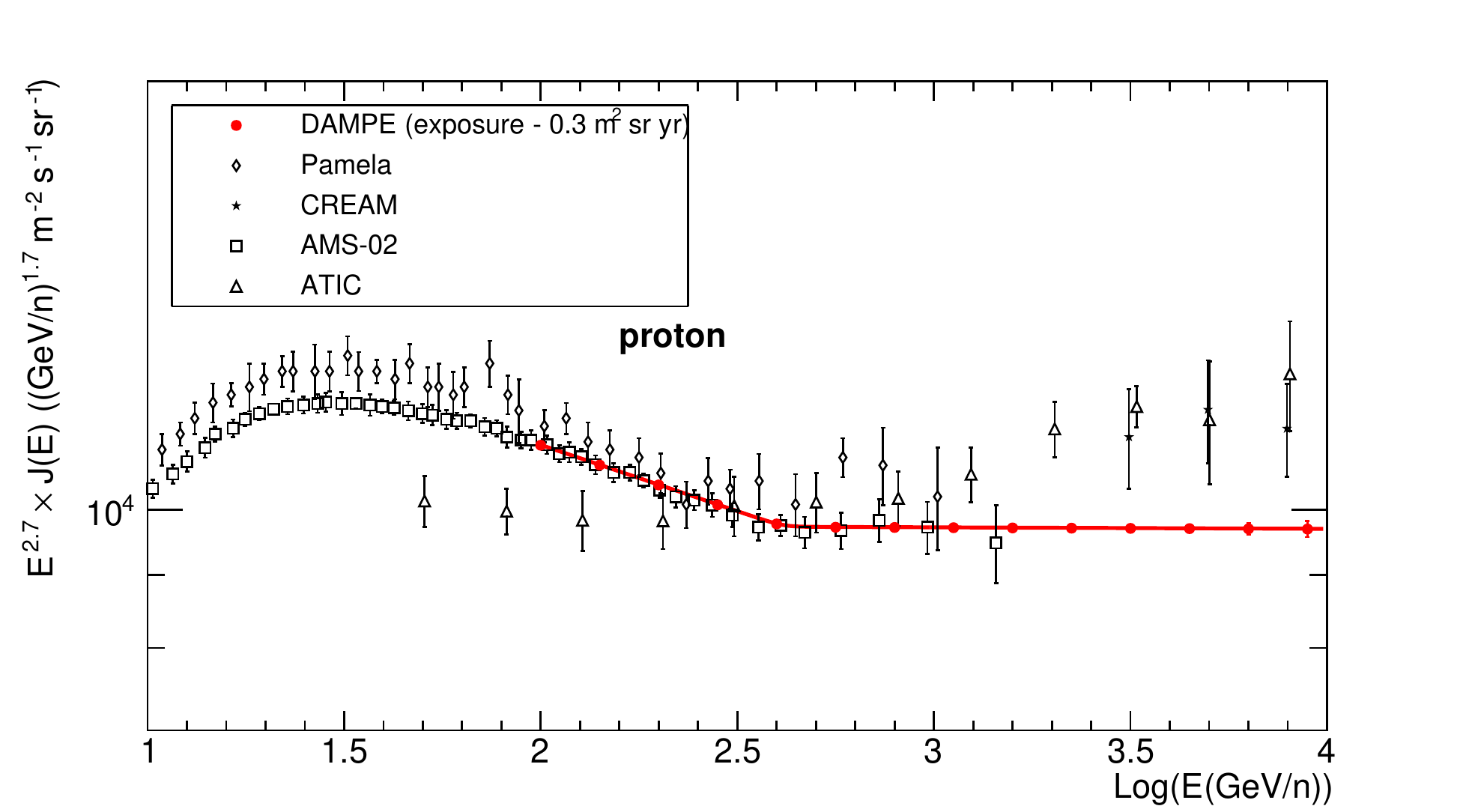}
\includegraphics[width=0.49\textwidth]{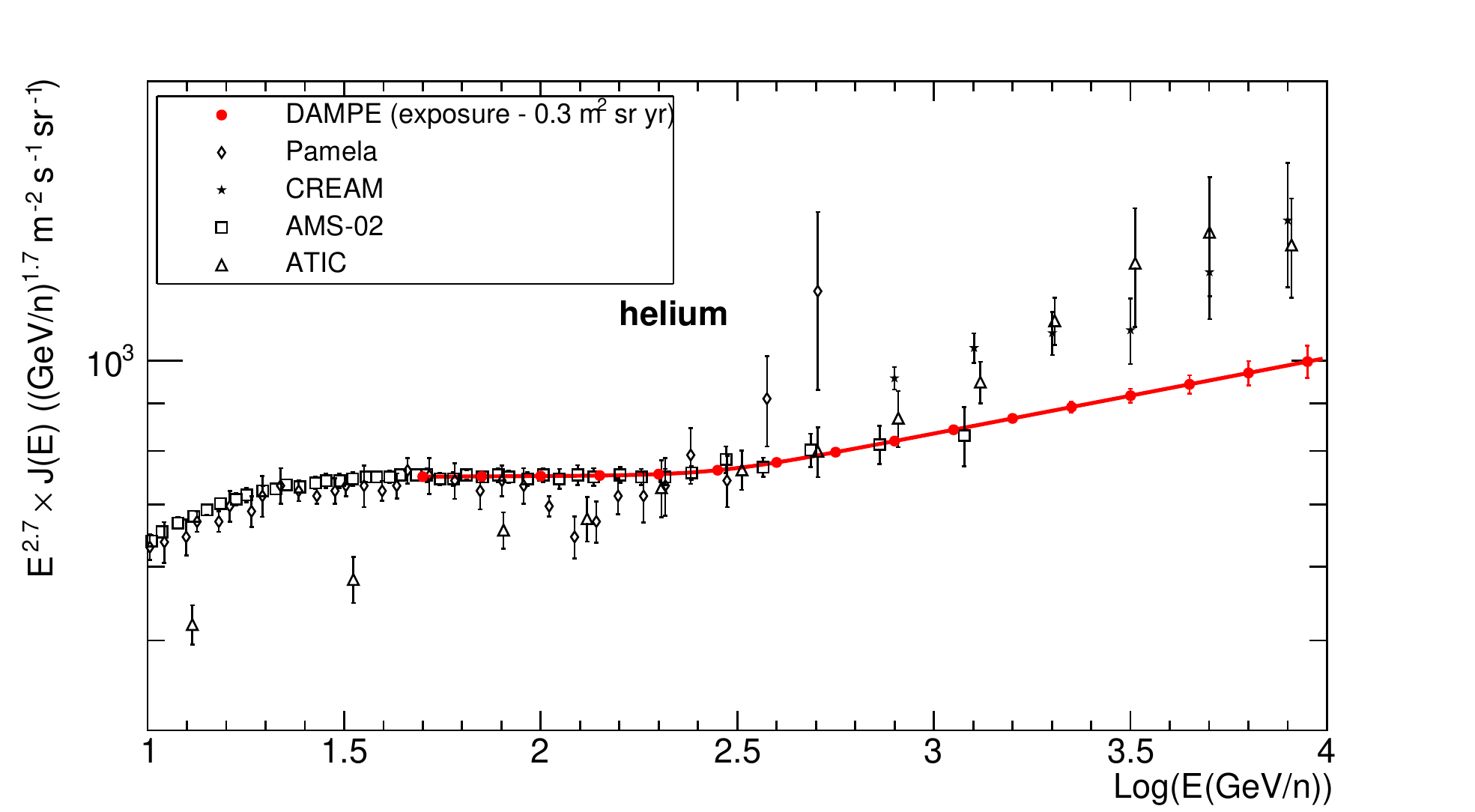}
\caption{Expected spectra of protons (left) and helium (right) that can be obtained by DAMPE, assuming the AMS-02 fluxes and their extrapolations,
with an exposure of 0.3 m$^2$ sr yr, compared with current measurements~\cite{Wefel2008,Ahn2010,Adriani2011,AMS2015,AMS2015b}.}
\label{fig:p_he}
\end{figure*}

\begin{figure}[!ht]
\centering
\includegraphics[scale=0.6]{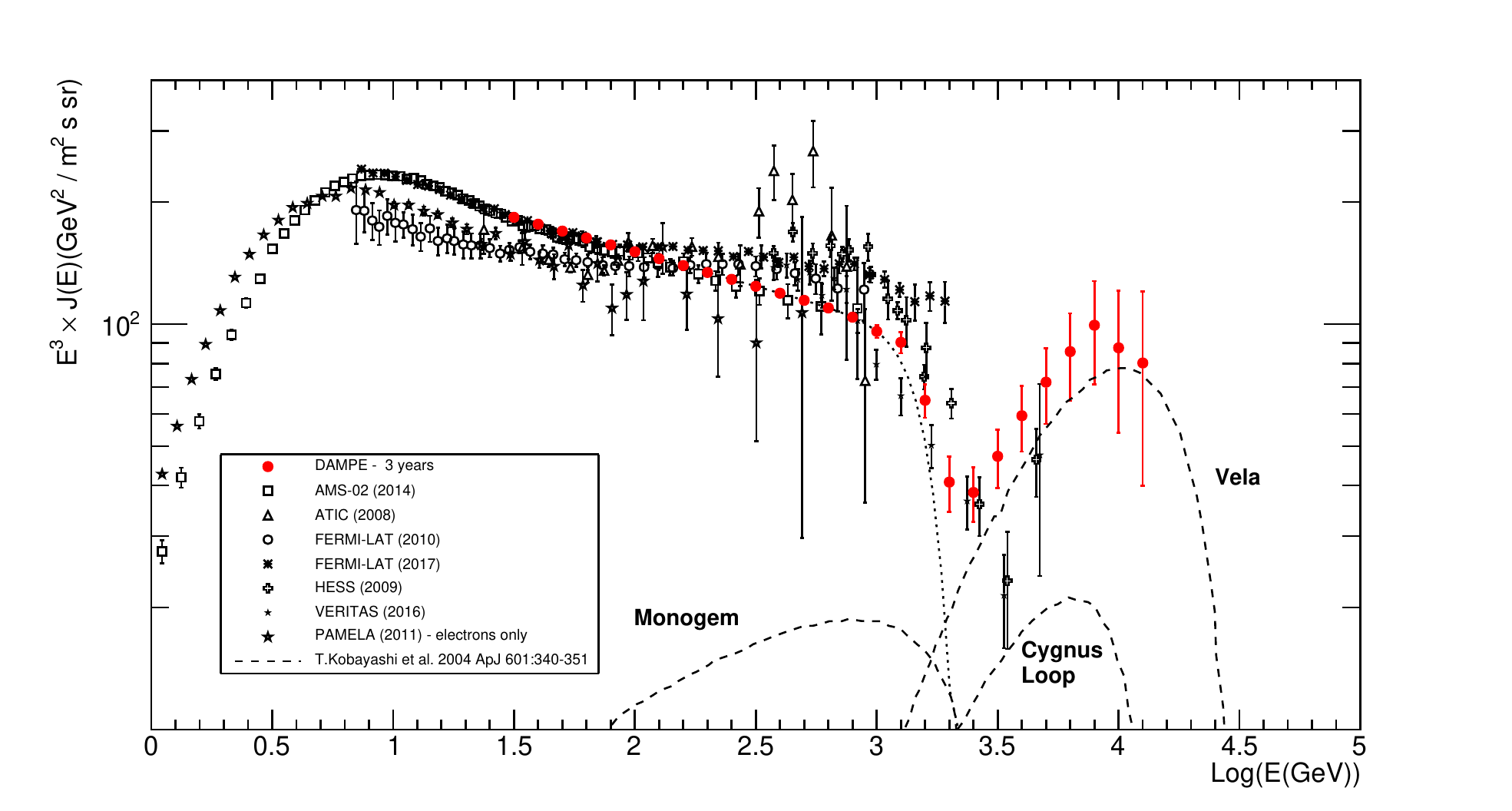}
\caption{Current measurement~\cite{Chang2008,Ackermann:2010ij,HESS2009,Adriani:2011xv,Fermi2017} and the expected spectrum of cosmic ray electrons (and positrons)
for three years operation of DAMPE, assuming the AMS-02 intensity, a cut-off and the contribution of Vela as calculated in~\cite{grasso2009}.
Note that some nearby young/middle-aged supernova remnants may give rise to additional TeV bump(s) in the spectrum.}
\label{fig:electrons-expected}	
\end{figure}

\begin{itemize}
\item The proton and helium are the most abundant components of cosmic rays.
The standard paradigm for particle acceleration and propagation predicts
single power-law spectra up to the so-called ``knee'' at $\sim 10^{15}$ eV.
Surprisingly, the spectra of cosmic-ray nuclei measured by ATIC
\cite{Wefel2008}, CREAM \cite{Ahn2010}, PAMELA \cite{Adriani2011} and
AMS-02 \cite{AMS2015,AMS2015b} all showed remarkable hardening at the
magnetic rigidity of several hundred GV. Such a result triggered various
modifications of the standard, simple picture of Galactic cosmic rays. Interesting
possibilities include the superposition of injection spectra
of the ensemble of sources \cite{Zatsepin2006,Yuan2011}, the effect of
local source(s) \cite{Vladimirov2012,Thoudam2012,Bi2017},  the complicated
acceleration of particles \cite{Ptuskin2013,Ohira2016}, or a non-uniform
diffusion coefficient \cite{Tomassetti2012,Guo2016}. Current spectral
measurements are, however, uncertain for energies above TeV/n. DAMPE will
be able to clearly measure the spectral changes and precisely determine
the high energy spectral indices of various nuclei species, as shown in
Fig. \ref{fig:p_he} for proton and helium. Furthermore, the DAMPE data will be able to test whether there are additional structures on the high energy cosmic-ray
spectra, as may be expected from nearby sources \cite{Bernard2013}. { Recently, the proton and helium spectra from the CREAM-III flight have been published and tentative breaks at $\sim 10-20$ TeV are displayed \cite{Yoon2017}. With the energy resolution of $\sim 20\%$ at such energies (see Fig.\ref{EneRes_Proton}; which is better than that of CREAM-III) and an expected exposure of $\sim 0.3{\rm m}^{2}~{\rm sr~yr}$, DAMPE will reliably test such a possibility.} The
cosmic-ray spectra up to 100 TeV by DAMPE will overlap with that measured
by the ground-based air shower experiments (e.g. \cite{ARGO2016}), which
can provide us with a full picture of the cosmic-ray spectra up to above
the knee. {\bf DAMPE can also measure the Boron-to-Carbon ratio, to about
5 TeV/n, which can effectively constrain the propagation parameters.}

\item Electrons/positrons contribute $\sim 1\%$ of the total amount of
cosmic rays. Unlike the nuclei, electrons/positrons lose their energies
efficiently during the propagation in the Galaxy. This is particularly true
for $\sim$TeV electrons/positrons which are expected to reach the
Earth only if the source is relatively nearby ($\lesssim 1$ kpc) and young
($\lesssim 10^5$ yr)~\cite{Fan2010,Ackermann:2010ip}. With an acceptance of $\sim
0.3~{\rm m^{2}~sr}$ at TeV energies (see Fig.\ref{fig:Acceptance}), DAMPE will precisely measure the trans-TeV behavior of the energy spectra of electrons/positrons, and determine the spectral structures e.g. spectral cut-off \cite{Chang2008,PAMELA2009,HESS2009,FermiLAT2009,Ackermann:2010ij,AMS2013,AMS2014a,AMS2014b}. As a consequence, DAMPE will be able to directly test a long-standing hypothesis that nearby pulsars or SNRs (e.g., Vela) are
efficient TeV electron accelerators \cite{Shen1970,Li2015PLB} by measuring
the spectrum and/or the spatial anisotropy of TeV electrons/positrons (see Fig.\ref{fig:electrons-expected} for an illustration).

\item DAMPE can also measure gamma-rays from Galactic and extra-galactic cosmic ray
accelerators such as SNRs, pulsars, quasars \cite{XuZL2017} etc. Although the effective acceptance of DAMPE is smaller than that of Fermi-LAT, DAMPE may play an auxiliary role in deep observations of these sources, especially in connection with ground-based measurements at hundreds of GeV.

\end{itemize}

\subsection{Probing the nature of dark matter}
As early as the 1930s, it was recognized that some matter in the Universe
is invisible~\cite{Zwicky1931}. The existence of this so-called dark
matter was gradually and firmly established since the early 1970s~\cite{Peacock1999}.
In the standard model of cosmology, the ordinary matter,
dark matter and dark energy share 4.9\%, 26.6\% and 68.5\% of today's
total energy density of the Universe. Compelling evidence shows that
the commonly existing dark matter is non-baryonic; however, the physical
nature of the dark matter particle is still poorly known~\cite{Bertone2005,Feng2010}.
Many theoretical models have been proposed,
and the suggested candidates span over a wide range of masses, mechanisms, and interaction
strengths~\cite{Bertone2005,Feng2010}. Among various candidates of dark
matter particles, one of the most attractive models is the weakly interacting
massive particle (WIMP), which is widely predicted in extensions of
the standard model of particle physics. The annihilation or decay of
WIMPs can give electromagnetic signals, primarily in the gamma-ray
band, as well as standard model particle products such as electrons/positrons,
neutrinos/anti-neutrinos and protons/antiprotons~\cite{Bertone2005,Feng2010,Turner1990,Gaskins2016}.

Anomalous peaks or structures in the energy spectra of cosmic rays (in
particular for electrons/positrons and antiprotons) and/or gamma rays from particular directions with accumulated dark matter distribution could indicate the existence of dark matter particles. In the past few years, several anomalous excesses had been reported in different cosmic-ray and
gamma-ray data sets, including the electron/positron excesses
~\cite{Chang2008,PAMELA2009,HESS2009,FermiLAT2009,Ackermann:2010ij,AMS2013,AMS2014a,AMS2014b},
the Galactic center GeV excess~\cite{Hooper2011,Daylan2016,Zhou2015,
Calore2015,Ajello2016}, the possible excesses in a few dwarf galaxies
~\cite{Geringer2015,Li2016PRD}, and the tentative $\sim 130$ GeV
gamma-ray line~\cite{Bringmann2012,Albert2015}. Recently, another
line-like structure around $43$ GeV from a number of galaxy clusters
was reported with the Fermi-LAT Pass 8 data~\cite{Liang2016}.
These candidate signals are either too weak to be claimed as a firm
detection, or can be interpreted with astrophysical models or potential
instrument systematics (e.g.,~\cite{Serpico2012,Bi2013,Finkbeiner2013,Gaskins2016}).

With its much improved energy resolution (see Fig.~\ref{fig:P-Energyresolution}),
DAMPE is suitable for the search
of gamma-ray line emission which can be expected in the annihilation
channel of $\gamma X$, where $X=(\gamma, Z_{0}, H)$ or other new neutral
particle. The energies of the monochromatic gamma-rays are given by
$E_\gamma=m_\chi [1-m_{X}^{2}/4m_\chi^{2}]$, where $m_\chi$ is the mass of dark matter particle~\cite{Bringmann2012}.
The firm detection of gamma-ray line(s) is believed to be a smoking-gun
signature of new physics, because no known astrophysical process is expected
to be able to produce such spectral feature(s). The high
resolution is also crucial to identify multiple lines with energies
close to each other~\cite{Su2012,Li2012}. A set of gamma-ray lines
would further provide convincing evidence of dark matter particles, and
could provide more information of physical properties of dark matter particle, such as
their couplings with standard model particles. Theoretically the line
emission is typically suppressed due to particle interactions through loop process, other scenarios e.g. the internal bremsstrahlung from dark matter annihilating into a pair of charged particles, might dominate the potential line signal~\cite{Bringmann2011}. Axions or axion-like particles (ALPs), if produced non-thermally, could be candidate of cold dark matter \cite{Preskill1983,Abbott1983,Dine1983}, which produce spectral fine structures due to the photon-ALP oscillation~\cite{Sikivie1983,Hooper2007,Ajello2016b}. DAMPE
will enhance our capability to search for monochromatic and/or sharp spectral structures of gamma-rays in the GeV-TeV range.

\begin{figure*}[!htbp]
\centering
\includegraphics[scale=0.5]{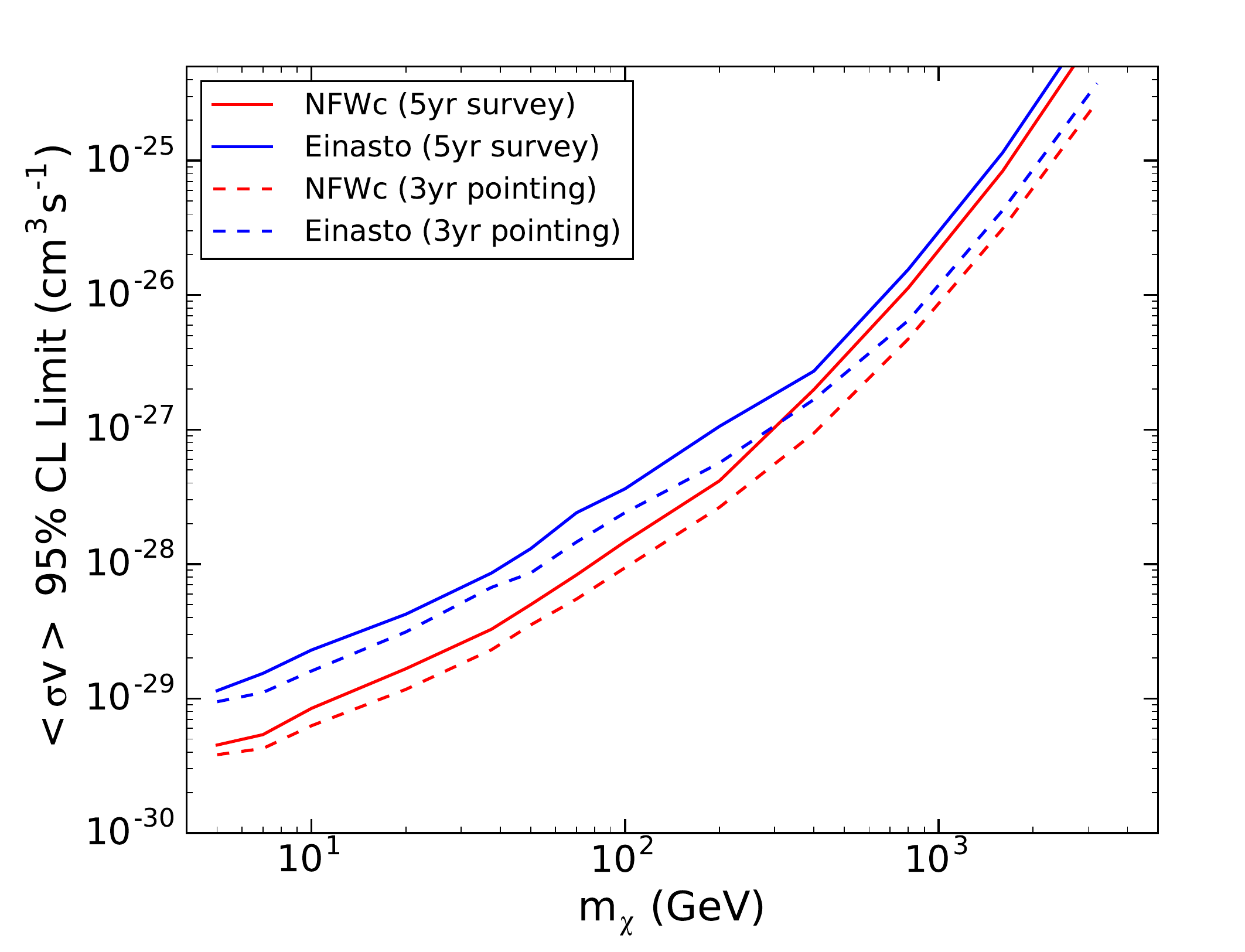}
\caption{Expected sensitivity of the gamma-ray line search by DAMPE in 3 years of targeted observations of the Galactic center,
and in 5 years of sky-survey observations.}
\label{fig:Dampe-line}
\end{figure*}

For illustration purpose, we take into account two types of dark matter density profile, including a
contracted Navarro-Frenk-White profile with $\gamma=1.3$ (NFWc)~\cite{NFW} and an Einasto profile
with $\alpha=0.17$~\cite{Einasto}. Following~\cite{Albert2015}, the Regions of Interest (ROIs) have
been taken as a 3$^\circ$ (16$^\circ$) circle centered on the Galactic center, respectively.
For the Einasto profile, we also mask the galactic plane region with $|l| > 6^\circ$ and $|b| < 5^\circ$.
The Galactic diffuse emission model {\it gll\_iem\_v06.fits}~\cite{Acero:2016qlg},
the isotropic diffuse model {\it iso\_P8R2\_SOURCE\_V6\_v06.txt}~\cite{fermifssc},
and 3FGL point sources of Fermi~\cite{Ackermann:2015yfk} have been combined to model the gamma-ray background.
The projected sensitivities of DAMPE in 3 years in case of targeted observations towards the Galactic center
and in 5 years of sky-survey observations are presented in Fig.~\ref{fig:Dampe-line}.

The electron/positron spectra can also be used to probe dark matter,
although the discrimination from local astrophysical sources may not be
trivial. In general the contribution from astrophysical sources is expected
to be non-universal, and may induce multiple features on the total energy spectra
~\cite{Malyshev2009}. DAMPE will accurately measure the energy spectra of electrons/positrons at trans-TeV energies to resolve these potential fine structures, which can be used to test/constrain dark matter models, in order to consistently explain the electron/positron excesses. With DAMPE's much improved energy resolution, possible new and fine spectral structures on the electron/positron spectra may be revealed as the existence of dark matter particles~\cite{Feng2013}.

\subsection{Studying high energy behaviors of gamma-ray transients and the diffuse emission}

DAMPE observes gamma-ray photons in the energy range of 10 GeV to TeV
and above with very high energy resolution.
Note that with the low energy trigger system the threshold can be as low
as $\sim 0.5$ GeV, with a reduced sampling rate by a factor of $\sim 64$.
Compared with Fermi-LAT, DAMPE has a smaller effective area and a higher
energy threshold. Therefore for stable GeV sources, DAMPE is not expected
to be competitive compared with Fermi-LAT due to limited counting statistics.
However, DAMPE may play a complementary, and possibly crucial role in
catching bright GeV-TeV transients, as each gamma-ray detector
can only cover part of the sky at the same time.

The collected gamma-ray data can be used to study the violent physical
processes behind activities of Active Galactic Nuclei (for instance
Mrk 421, 3C279 and 3C454.3), the Crab flares, and some bright gamma-ray
bursts (GRBs) such as GRB 130427A~\cite{Ackermann2014,Fan2013}.
Bright short GRBs with an isotropic GeV $\gamma$-ray energy release of $\geq 2\times 10^{51}$ erg, if taking place within $\sim 400$ Mpc,
might be detectable by DAMPE and could also serve as the electromagnetic counterparts~\cite{Abbott:2016gcq,Abbott:2016iqz}
of advanced LIGO/Virgo gravitational wave events~\cite{Abbott2016} or IceCube PeV neutrino events~\cite{Aartsen2013}. High energy gamma-ray observations can also be used to probe the extragalactic background light, the intergalactic magnetic field, and the fundamental
physics such as Lorentz invariance violation and quantum gravity.

The hadronic interaction of cosmic ray nuclei with the interstellar medium produces bright diffuse gamma-ray emission, primarily along the Galactic plane. In some regions (e.g. the Galactic center ridge and the Cygnus region) of the Galactic plane, fresh cosmic ray accelerators may light up surrounding materials with gamma-ray emission on top of the diffuse background~\cite{Aharonian2006,Abdo2007}. Thanks to the improved hadron
rejection power of DAMPE ($>10^5$), it is possible to measure the diffuse
gamma-ray emission up to TeV energies without significant contamination
from cosmic rays. DAMPE has the potential to reliably detect $>1$ TeV gamma-rays
in space. The DAMPE mission will provide a crucial overlap energy range to bridge the space and
ground-based gamma-ray measurements, providing effective constraints
on the origin of cosmic rays in the Milky Way.

\section{Summary}
DAMPE was successfully launched into a sun-synchronous orbit at the altitude of 500 km on December $17^{\rm th}$ 2015.
The combination of the wide field of view, the large effective area and acceptance and the excellent energy resolution offers new opportunities for advancing
our knowledge of cosmic rays, dark matter and high energy astronomy.\\

{\it Acknowledgments:}  The DAMPE mission was funded by the strategic
priority science and technology projects in space science of the Chinese
Academy of Sciences (No. XDA04040000 and No. XDA04040400). In China this
work is also supported in part by National Key Research and Development
Program of China (No. 2016YFA0400200), the National Basic Research Program
(No. 2013CB837000), National Natural Science Foundation of China under
grants No. 11525313 (i.e., Funds for Distinguished Young Scholars),
No. 11622327 (i.e., Funds for Excellent Young Scholars), No. 11273070,
No. 11303096, No. 11303105, No. 11303106, No. 11303107, No. 11673075,
U1531126, U1631111 and the 100 Talents program of Chinese Academy of Sciences.
In Europe the work is supported by the Swiss National Science Science
Foundation, the University of Geneva, the Italian National Institute for
Nuclear Physics, and the Italian University and Research Ministry.
We also would like to take this opportunity to thank the scientific
laboratories and test facilities in China and Europe (in particular CERN
for provision of accelerator beams) that assisted the DAMPE team during
the qualification phases.


\begin{table}[!ht]
\centering
\caption{Abbreviations in the main text.} \label{tab:abbre}
\begin{tabular}{ll}
\hline
Abbreviation &  Full expression   \\
\hline
3FGL         &  3rd Fermi Gamma-ray LAT catalog \\
ACE          &  Advanced Composition Explorer \\
ADC          &  Analog-Digital Conversion \\
AGILE        &  Astro‐rivelatore Gamma a Immagini LEggero \\
ALP          &  Axion-Like Particle \\
AMS          &  Alpha Magnetic Spectrometer \\
ASDC         &  ASI Science Data Center \\
ASIC         &  Application Specific Integrated Circuit \\
ATIC         &  Advanced Thin Ionization Calorimeter \\
BESS         &  Balloon-borne Experiment with Superconducting Spectrometer \\
BGO          &  Bismuth Germanium Oxide \\
CALET        &  CALorimetric Electron Telescope \\
CAPRICE      &  Cosmic AntiParticle Ring Imaging Cherenkov Experiment \\
CERN         &  European Organization for Nuclear Research \\
CFRP         &  Carbon Fiber Reinforced Plastics \\
CPU          &  Central Processing Unit \\
CRC          &  Cyclic Redundancy Check \\
CREAM        &  Cosmic Ray Energetics And Mass \\
DAMPE        &  DArk Matter Particle Explorer \\
DAMPESW      &  DAMPE Software Framework \\
DAQ          &  Data AcQuisition  \\
EGRET        &  Energetic Gamma Ray Experiment Telescope \\
EJ           &  Eljen \\
EQM          &  Engineering Qualification Model \\
FEE          &  Front-End Electronics \\
Fermi-LAT    &  Fermi Large Area Telescope \\
FoV          &  Field of View \\
FPGA         &  Field-Programmable Gate Array \\
GC           &  Gating Circuit \\
GEANT        &  GEometry ANd Tracking \\
GPS          &  Global Positioning System \\
GRB          &  Gamma-Ray Burst \\
GSS          &  Ground Support System \\
H.E.S.S.     &  High Energy Stereoscopic System \\
HAWC         &  High-Altitude Water Cherenkov \\
HEAO         &  High Energy Astronomy Observatory \\
HEAT         &  High Energy Antimatter Telescope \\
IMAX         &  Isotope Matter Antimatter eXperiment \\
IMP          &  Interplanetary Monitoring Platform \\
INFN         &  Istituto Nazionale di Fisica Nucleare \\
LIGO         &  Laser Interferometer Gravitational wave Observatory \\
\hline
\end{tabular}
\end{table}

\begin{table}[!ht]
\centering
\centerline{Table 6 --- continued.}
\begin{tabular}{ll}
\hline
Abbreviation &  Full expression   \\
\hline
MAGIC        &  Major Atmospheric Gamma Imaging Cherenkov Telescopes \\
MC           &  Monte Carlo \\
MIP          &  Minimum Ionizing Particle \\
MPV          &  Most Probable Value \\
NFW          &  Navarro-Frenk-White \\
NUD          &  NeUtron Detector \\
PAMELA       &  Payload for Antimatter Matter Exploration and Light-nuclei Astrophysics \\
PAO          &  Pierre Auger Observatory \\
PDPU         &  Payload Data Process Unit \\
PHC          &  Peak Holding Chip \\
PMT          &  PhotoMultiplier Tube \\
PMU          &  Payload Management Unit \\
PSD          &  Plastic Scintillator strip Detector \\
ROI          &  Region of Interest \\
SAS          &  Scientific Application System \\
SC           &  Shaping Circuit \\
SPS          &  Super Proton Synchrotron \\
SSD          &  Silicon micro-Strip Detector \\
STK          &  Silicon-Tungsten tracKer-converter \\
TFH          &  Tracker Front-end Hybrid \\
TMVA         &  Toolkit for MultiVariate data Analysis \\
TRB          &  Tracker Readout Board \\
VERITAS      &  Very Energetic Radiation Imaging Telescope Array System \\
WIMP         &  Weakly Interacting Massive Particle \\
\hline
\end{tabular}
\end{table}

\end{document}